\documentclass[preprint,12pt,authoryear]{elsarticle}
\usepackage{amssymb,amsfonts,amsthm}
\usepackage{geometry}
\usepackage{graphics,subfigure}
\usepackage{hyperref}
\usepackage{pifont}
\usepackage{natbib}
\usepackage{textcomp}
\usepackage{gensymb}
\usepackage[fleqn]{amsmath}
\usepackage{color}
\usepackage{lineno}
\usepackage{bm}
\numberwithin{equation}{section}

\begin{document}
\begin{frontmatter}
\title{Post-buckling of fiber-reinforced soft tissues}

\author[add1,add2,add3]{Yang Liu\fnref{fn1}\corref{cor1}}
\ead{liuy3@maths.ox.ac.uk, tracy\_liu@tju.edu.cn} 
\author[add1]{Rui-Cheng Liu\fnref{fn1}}
\author[add1]{Wanyu Ma\fnref{fn1}}
\author[add2]{Alain Goriely\fnref{fn1}}
\address[add1]{Department of Mechanics, School of Mechanical Engineering, Tianjin University, Tianjin 300350, China}
\address[add2]{Mathematical Institute, University of Oxford, Oxford, OX2 6GG, UK}
\address[add3]{National Key Laboratory of Vehicle Power System, Tianjin 300350, China}

\cortext[cor1]{Corresponding author.}
\fntext[fn1]{All authors contributed equally to this work.}
    
\begin{abstract}
Fiber-reinforcement is  a universal feature of many  biological tissues. It involves the interplay between fiber stiffness, fiber orientation, and the elastic properties of the matrix, influencing pattern formation and evolution in layered tissues. Here, we investigate the deformation of a compressed film bonded to a half-space, where either the film or the substrate exhibits anisotropy. Within the framework of finite elasticity, we formulate nonlinear incremental equations, enabling linear and weakly nonlinear analyses. These analyses yield exact bifurcation conditions and an amplitude equation for surface wrinkling. In particular, for a simple fiber-reinforced model, we show that the bifurcation can be supercritical or subcritical depending on the ratio between the substrate and the film moduli. These findings underscore the pivotal role of fiber-reinforcement in shaping pattern formation in anisotropic tissues and provide insights into the morphological evolution of biological tissues.
\end{abstract}

\end{frontmatter}

\section{Introduction} 
Bilayered structures, where a thin but stiff film adheres to a thick and compliant substrate, are widely observed in nature and engineering applications  \citep{Genzer2006SM,Li2012SM,Goriely2017}.  This type of structure is often susceptible to wrinkling formation, which  can be induced through compressive stresses, generated from either compression \citep{Chen2004JAM,Kim2011NM}, thermal expansion \citep{Barnes2023PRL}, growth \citep{Li2011JMPS}, or swelling \citep{Liu2022MMS}. For instance, it is well appreciated that human skins initiate wrinkles with aging \citep{Zhao2020JMBBM,Chavoshnejad2021,Lynch2022SR}, brain develops complex folds due to the fast expansion of the cortical layer \citep{Budday2015PM,Holland2020EPJ,Wang2021BMM}, and fruits and vegetables form wrinkles because of differential growth between pericarp and sarcocarp \citep{Yin2009JMPS,Dai2014EPL}.

Both theoretically and experimentally, a film/substrate bilayer  serves as a foundational model for investigating complex pattern formation in layered structures. Consequently, significant effort has been devoted to determining the critical conditions for surface wrinkling by solving the eigenvalue problem associated with the linearized governing equations \citep{Shield1994JAM,Bigoni1997IJSS,Cai2000IJSS,Liu2014IJES,Zhang2017JAM,Alawiye2019PTRSA,Liu2024JMPS}. This approach allows the simultaneous determination of critical loads and the corresponding wrinkled patterns. However, post-bifurcation behavior can only be analyzed within the nonlinear regime. In the context of finite elasticity, this analysis is commonly conducted by superimposing an incremental deformation onto a finitely deformed state, a method now referred to as incremental theory \citep{Biot1965,Ogden1984,Goriely2017}. Following this original analysis, \citet{Cai1999PRSA}  derived an amplitude equation for surface wrinkling of a neo-Hookean film bonded to another neo-Hookean half-space and identified a critical modulus ratio of two layers where a transition between supercritical and subcritical bifurcations takes place. Later, they  investigated the effect of compressibility and secondary period-doubling bifurcations \citep{Cai2019IJNM,Fu2015SIAM}. \cite{Hutchinson2013} re-discovered the same transition shown in \cite{Cai1999PRSA} by including pre-stretch effects. For three-dimensional wrinkling, \citet{Ciarletta2014JMPS} performed a weakly nonlinear analysis for soft layers with equi-biaxial strain and analytically studied the wrinkle to fold transition. Using the finite growth theory \citep{BenAmar2005JMPS}, \citet{Ciarletta2015IJNM} developed a semi-analytical method for a growing layer, and later \citet{Jin2019IJSS} extended the nonlinear analysis to tubular tissues to trace the post-buckling evolution. Recently, \cite{Alawiye2020JMPS} performed the post-buckling analysis assuming that the compressive stress can be generated by either compression or the growth of thin film. 

The aforementioned studies reveal that when the film and substrate possess similar elastic moduli, the bifurcation transitions to a subcritical nature. Analyzing the post-buckling behavior in cases of subcritical bifurcation presents significant theoretical challenges, as multiple modes often interact near the critical point, complicating the analysis. \citet{Fu2015PRSA} employed a Fourier expansion to derive nonlinear amplitude equations for subcritical bifurcation, revealing that a localized solution can emerge as the limit of modulated periodic solutions. More recently, \citet{Shen2024JMPS} examined a similar problem by considering growth effects and numerically identified distinct characteristic domains where different patterns exist. 

Wrinkles in layered structures continue to attract significant research interest across diverse fields due to their widespread applications in flexible electronics, sensors, and related technologies \citep{Sun2006NN,Cui2016IEEE,Lee2022Small}, surface engineering \citep{Yang2010AFM,Zhang2012JACS,Li2017AOM,Lee2021EML}, elastic metamaterials \citep{Liu2022EML,Han2024LPR}, and bio-inspired design \citep{Tan2020NML,Zhao2024SI}. As a result, many different features of the problem have been studied, including the influence of viscosity \citep{Huang2005JMPS,Liu2022IJSS}, multi-layering \citep{Cheng2014IJSS,Zhou2022MMS}, growth or swelling \citep{Budday2014JMPS,Zhao2015JMPS,Eskandari2016,Polukhov2023JMPS}, grading modulus \citep{Chen2017PRSA,Liu2020IJNLM},  curvature \citep{Jia2018PRE,Liu2025Nonlinearity} and pattern selection in three {\color{black}dimensions} \citep{Cheewaruangroj2019SM}.

This paper focuses on anisotropic bilayers. In engineering applications, anisotropy can be used to tune pattern selection \citep{Im2008JMPS,Song2010APL, Yin2018IJSS}. Similarly, many biological tissues, such as skin and arteries, are naturally endowed with fibers, making them intrinsically anisotropic \citep{Chavoshnejad2021, Hill2012JB}. Consequently, it is of fundamental importance to develop mechanical models that address buckling instabilities and post-buckling evolution, providing insights into the formation and progression of various morphological patterns.

\citet{Im2008JMPS} studied the wrinkled pattern of anisotropic films resting on viscoelastic substrates. Based on the F\"{o}ppl-von K\'{a}rm\'{a}n plate theory, \citet{Yin2018IJSS} explored analytically and numerically  pattern formation and evolution in an orthotropic film bonded to a compliant substrate. Using linear elasticity, \citet{Guo2023} investigated the influence of fiber orientation on surface wrinkling in anisotropic substrates. However, it has been noted that the surface instability of many biological tissues is often associated with highly nonlinear constitutive responses \citep{Pocivavsek2018NP}, highlighting the necessity of addressing these phenomena within the framework of finite elasticity.

To account for fiber-reinforcement, one of the most widely used models is the Holzapfel-Gasser-Ogden (HGO) model, which was developed for arteries \citep{Holzapfel2000JE,Gasser2006JRSI,Holzapfel2010PRSA}. The HGO model is also applicable to various biological tissues exhibiting collagen fiber distributions \citep{Holzapfel2019PRSA}. Notably, a key assumption of the HGO model is that fibers are nearly inextensible and incapable of bearing compression. To address limitations associated with fiber extensibility, \citet{Horgan2005JMPS}, inspired by the Gent model for natural rubber \citep{Gent1996}, proposed a constitutive theory for fiber-reinforced incompressible solids, accommodating fibers with finite extensibility. Building on these advancements, \citet{Stewart2016EML} explored brain morphogenesis by analyzing pattern evolution in a film bonded to a fiber-reinforced substrate under compression. They found that the folding pattern is nearly independent of the fiber orientation. \citet{Nguyen2020BMM} conducted a detailed analysis of the effects of fiber orientation, fiber stiffness, and dispersion on the onset of surface wrinkling in a neo-Hookean film bonded to an HGO substrate. Additionally, they employed finite element simulations to investigate the post-buckling evolution. {\color{black} More recently, \citet{Mirandola2023JAP} derived a scaling law to characterize the critical wrinkled state in anisotropic bilayers where the substrate is reinforced by two families of fibers. \citet{Altun2023MoSM} numerically explored growth-induced surface wrinkling when the film has a single family of fibers, using a three-dimensional framework based on a five-field Hu–Washizu-type mixed variational formulation.}

We emphasize that all the aforementioned studies on the post-buckling behavior of anisotropic film/substrate bilayers have primarily relied on finite element simulations. Analytical solutions for post-buckling behavior remain scarce, and the bifurcation characteristics of the initial instability, particularly the role of anisotropy, are not yet fully understood. This paper aims to address these gaps by performing an analytical post-buckling  analysis. Additionally, we seek to develop a unified framework capable of analyzing cases where either the thin film or the substrate is anisotropic. This approach will enable a systematic and qualitative comparison between the two scenarios, providing  insights into the influence of anisotropy on instability and pattern formation.

The remainder of this paper is organized as follows. Section \ref{problem-formulation} presents a summary of the  fundamental equations. In Section \ref{linear-bifurcation-analysis}, we conduct a linear bifurcation analysis to determine the exact bifurcation condition. The effects of fiber orientation and stiffness on the critical state are examined, and mode transitions are discussed. Section \ref{nonlinear-analysis} focuses on a weakly nonlinear analysis, where an analytical amplitude equation is derived. The transition between supercritical and subcritical bifurcations is identified, and the post-buckling evolution is analyzed in detail. {\color{black}We provide several comparisons with existing studies and outline potential directions for future work in Section \ref{discussions}.} Finally, conclusions are provided in Section \ref{conclusion}.

\section{Problem formulation and incremental theory}\label{problem-formulation}
 \begin{figure}
  \centering
  \includegraphics[width=1\textwidth]{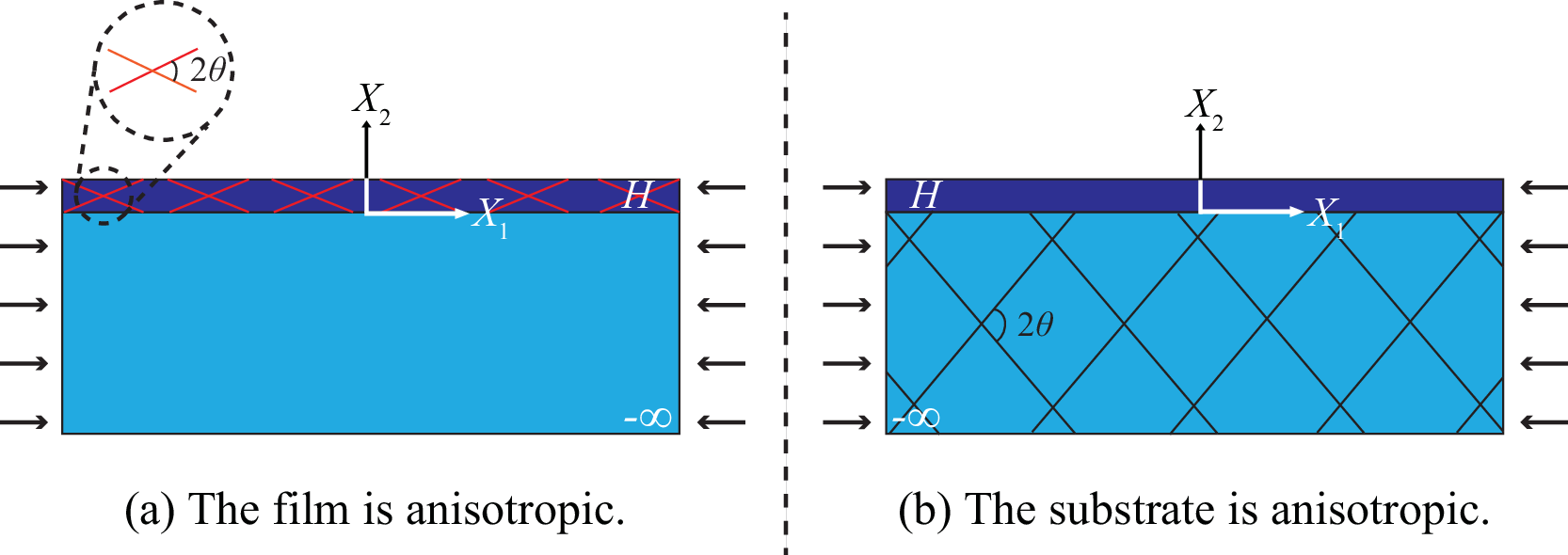}
  \caption{Geometry of a bilayer in the reference configuration $\mathcal{B}_0$ where either the film or the substrate is composed of a fiber-reinforced anisotropic material.}
  \label{bilayer}
\end{figure}

Consider an elastic bilayer where either the film or the substrate is an anisotropic material and both incompressible layers are subject to the same axial compression as shown in Figure \ref{bilayer}. Initially,  the film has a thickness $H$, and the substrate is a half-space. Anisotropy is introduced by two families of fibers oriented at an angle $\pm\theta$ with respect to $X_1$, the horizontal axis. We  use the Cartesian coordinate system and adopt a plane-strain condition in the $(X_1,X_2)$-plane. 

The bilayer is subject to a compressive stress along the $X_1$-direction, leading for small values of the compression to a homogeneous deformation occurs in which the thickness of the film increases to $h$. For larger values, we expect that an instability will take place and the homogeneous will be replaced by a surface wrinkling profile at a critical compressive strain. Here, we  focus on the post-buckling behavior by performing a weakly nonlinear analysis using incremental theory \citep{Ogden1984,Fu1999CMT}. We begin by formulating the nonlinear incremental equations for a general hyperelastic material. 

We use $\mathbf{X}=(X_1,X_2)$ and $\mathbf{x}=(x_1,x_2)$ to represent the position vectors of a material point in the stress-free reference configuration $\mathcal{B}_0$ and in the homogeneously stressed configuration $\mathcal{B}_r$, respectively.  Subsequently, we superimpose a small-amplitude deformation on {\color{black}$\mathcal{B}_r$}, leading to the final configuration denoted by $\mathcal{B}_t$, with associated position vector  $\tilde{\mathbf{x}}=(\tilde{x}_1,\tilde{x}_2)$ that can be written as
\begin{equation}
    \tilde{\mathbf{x}}=\mathbf{x}+\mathbf{u}(\mathbf{x}),
    \label{relation}
\end{equation}
where $\mathbf{u}$ is the incremental displacement from $\mathcal{B}_r$ to $\mathcal{B}_t$. Introducing the common orthonormal basis in all three configurations $\{\mathbf{e}_1,\mathbf{e}_2\}$, we have $\mathbf{u}=u_1\mathbf{e}_1+u_2\mathbf{e}_2$.

The deformation gradients $\mathbf{F}$ from {\color{black}$\mathcal{B}_0\to \mathcal{B}_t$} and $\bar{\mathbf{F}}$ from {\color{black}$\mathcal{B}_0\to \mathcal{B}_r$} are given by 
\begin{equation}
   {F}_{i A}=\frac{\partial \tilde{x}_i}{\partial X_A},\quad {\bar{F}}_{i A}=\frac{\partial x_i}{\partial X_A}. 
\end{equation}
They are related by
\begin{equation}
   F_{iA}=\left(\delta_{ij}+u_{i,j}\right)\bar{F}_{jA},
   \label{deformation-relation}
\end{equation}
where $\delta_{ij}$ is the usual Kronecker delta and we define a lower case $j$ behind a comma to be the derivative with respect to $x_j$ while a capital $A$ letter would indicate the derivative with respect to $X_A$.

The incompressibility constraint yields 
\begin{equation}
    \det\mathbf{F}=1,\quad \det\bar{\mathbf{F}}=1.
    \label{incom-con}
\end{equation}
The left and right Cauchy-Green deformation tensors read
\begin{equation}
\mathbf{B}=\mathbf{FF}^{\mathsf{T}}, \quad \mathbf{C}=\mathbf{F}^{\mathsf{T}}\mathbf{F},
\label{Cauchy-Green}
\end{equation}
with similar expressions in $\mathcal{B}_r$.

The first Piola-Kirchhoff stress tensors in $\mathcal{B}_r$ and $\mathcal{B}_t$ are given by
\begin{equation}
    \pi_{i A}=\frac{\partial W(\mathbf{F})}{\partial F_{i A}}-p F_{A i}^{-1},\quad \bar{\pi}_{i A}=\frac{\partial W(\bar{\mathbf{F}})}{\partial \bar{F}_{i A}}-\bar{p} \bar{F}_{A i}^{-1},
\end{equation}
where $W$ denotes the strain-energy density, {\color{black}$p$ and $\bar{p}$} are the Lagrange multipliers enforcing the incompressibility conditions in $\eqref{incom-con}$. 

In the absence of body forces, the equilibrium conditions for both states $\mathcal{B}_r$ and $\mathcal{B}_t$ are 
\begin{equation}\label{eq-motion}
\bar{\pi}_{i A, A}=0, \quad \pi_{i A, A}=0,
\end{equation}
where we adopt the  summation convention over repeated indices unless specified otherwise.

To address the stability of $\mathcal{B}_r$, we introduce the incremental stress tensor
\begin{equation}
    \bm\chi=\left(\bm\pi-\bar{\bm\pi}\right)\bar{\mathbf{F}}^\mathsf{T}, \quad \mathrm{or}\quad \chi_{i j}=\left(\pi_{i A}-\bar{\pi}_{i A}\right) \bar{F}_{j A}.
\label{incr-str-com}
\end{equation}

In view of \eqref{eq-motion}, \eqref{incr-str-com} and by virtue of the identity $\operatorname{div}\bar{\mathbf{F}}=\bm0$, we arrive at the incremental equilibrium equation 
\begin{equation}
    \label{eq-incr}
    \operatorname{div}\bm\chi^\mathsf{T}=\bm 0, \quad \mathrm{or}\quad \chi_{i j, j}=0.
\end{equation}
where `div'  is the divergence operator in $\mathcal{B}_r$.

For an infinitesimal incremental displacement $\mathbf{u}(\mathbf{x})$, we expand  $\bm\chi$ in the incremental displacement to obtain
\begin{align}
\notag\chi_{i j}&=\mathcal{A}_{j i l k}^1 u_{k, l}+\frac{1}{2} \mathcal{A}_{j i l k n m}^2 u_{k, l} u_{m, n}+\frac{1}{6} \mathcal{A}_{j i l k n m q p}^3 u_{k, l} u_{m, n} u_{p, q} \\
&+\bar{p}\left(u_{j, i}-u_{j, k} u_{k, i}+u_{j, k} u_{k, l} u_{l, i}\right)-p^*\left(\delta_{j i}-u_{j, i}+u_{j, k} u_{k, i}\right)+\mathcal{O}\left(|u_{i,j}|^4\right),
\label{exp-chi}
\end{align}
where $\mathcal{A}^1$, $\mathcal{A}^2$, $\mathcal{A}^3$ are, respectively, the first-, second-, third-order instantaneous elastic moduli and $p^*=p-\bar{p}$ is the pressure increment. In the above expression, we retain nonlinearities up to the third order, as required in a generic weakly nonlinear analysis \citep{Cai1999PRSA,Fu1999CMT,Jin2019IJSS,Alawiye2020JMPS}.

Explicitly, the incremental elastic moduli are \citep{Chadwick,Ogden1984,Fu1999CMT}.
\begin{equation}
\begin{aligned}
&\mathcal{A}_{j i l k}^1=\left.\bar{F}_{j A} \bar{F}_{l B} \frac{\partial^2 W}{\partial F_{i A} \partial F_{k B}}\right|_{\mathbf{F}=\bar{\mathbf{F}}},\\
&\mathcal{A}_{j i l k n m}^2=\left.\bar{F}_{j A} \bar{F}_{l B} \bar{F}_{n C}\frac{\partial^3 W}{\partial F_{i A} \partial F_{k B} \partial F_{m C}}\right|_{\mathbf{F}=\bar{\mathbf{F}}},\\
&\mathcal{A}_{j i l k n m q p}^3=\left.\bar{F}_{j A} \bar{F}_{l B}\bar{F}_{n C}\bar{F}_{q D} \frac{\partial^4 W}{\partial F_{i A} \partial F_{k B} \partial F_{m C}\partial F_{p D}}\right|_{\mathbf{F}=\bar{\mathbf{F}}}.
\end{aligned}\label{eq:instant moduli}
\end{equation}

On substituting \eqref{exp-chi} into (\ref{eq-incr}), we obtain the incremental equations up to third order:
\begin{equation}
\mathcal{A}_{j i l k}^1 u_{k, l j}+\mathcal{A}_{j i l k n m}^2 u_{m, n} u_{k, l j}+\frac{1}{2} \mathcal{A}_{j i l k n m q p}^3 u_{m, n} u_{p, q} u_{k, l j}-p_{, j}^*\left(\delta_{j i}-u_{j, i}+u_{j, m} u_{m, i}\right)+\mathcal{O}\left(|u_{i,j}|^4\right)=0.
\label{eq-third-order}
\end{equation}
In addition, the incompressibility condition $\eqref{incom-con}_1$ gives \citep{Fu1999CMT}
\begin{equation}
u_{i, i}=\frac{1}{2} u_{m, n} u_{n, m}-\frac{1}{2}\left(u_{i, i}\right)^2-\det\left(u_{m, n}\right).
\label{inc-nonlinear}
\end{equation}
Currently, we have formulated the general governing equations in \eqref{eq-third-order} and \eqref{inc-nonlinear} for further analysis. Note that we still need to give the incremental boundary conditions.

We from now on focus on the surface wrinkling in compressive bilayers where either the layer or the substrate may be {\color{black}anisotropic}, as shown in Figure \ref{bilayer}. To distinguish between two layers, we denote quantities associated with the half-space by using an over-hat, whereas quantities without a hat belong to the film. For quantities that take the same form in both layers, the hat is omitted if the quantity is evaluated in the domain of the half-space. For example, the deformation gradient of the homogeneous deformation is valid for both layers and we simply write
\begin{equation}
 \bar{\mathbf{F}} = \rm{diag} \left\{\lambda, \lambda^{-1}\right\},
 \label{def-gradient-hg}
\end{equation}
where $\lambda$ is the principal stretch in the $X_1$-direction and $1/\lambda$ is the principal stretch in the $X_2$-direction to satisfy the  incompressibility constraint. Therefore, the deformed thickness of the film becomes
\begin{equation}
    h=\dfrac{H}{\lambda}.
\end{equation}

For boundary conditions, we consider a traction-free top surface, perfect bonding at the interface, and zero displacement at infinity. In doing so, the incremental boundary conditions and continuity conditions are summarized as follows
\begin{equation} \label{inc-bc}
\left.\begin{aligned}
&\chi_{i2}=0, \quad \text { on } x_{2}=h,\\&
\hat{u}_{i}=u_{i}, \quad \hat{\chi}_{i2}=\chi_{i2}, \quad \text { on } x_{2}=0,\\&
\hat{u}_i\rightarrow0,\quad\text{ as } x_2\rightarrow-\infty,
\end{aligned}\right\}\quad i=1,2.
\end{equation}

Equations \eqref{eq-third-order}, \eqref{inc-nonlinear} and their counterparts for the half-space, together with \eqref{inc-bc}, formulate the incremental system governing wrinkling initiation and evolution in compressive bilayers. Later, we first perform a linear analysis and then a weakly nonlinear analysis.

\section{Linear bifurcation analysis}\label{linear-bifurcation-analysis}
We perform a linear bifurcation analysis to find the critical compressive strain. To this end, we use the linear part of the incremental system established in the previous section. To avoid redundant derivations, we  briefly summarize the general result of the eigenvalue problem stemming from the linearized versions of \eqref{eq-third-order} and \eqref{inc-nonlinear}, which are given by
\begin{equation} \label{eq-linear}
{\mathcal{A}}_{j i l k}^1 u_{k,l j}-{p}_{, i}^*=0,\quad {u}_{i, i}=0, \quad -\infty < x_2 \leqslant h.
\end{equation}
It is convenient to introduce a \textit{stream} function $\psi(x_1, x_2)$ such that
\begin{equation}\label{stream-function}
u_1=\psi_{, 2}, \quad u_2=-\psi_{, 1}.
\end{equation}
By substituting (\ref{stream-function}) into (\ref{eq-linear}) and then eliminating $p^*$ through cross-differentiation, we obtain the fourth-order equation
\begin{equation}
\label{eq-fourth-order-PDE}
\mathcal{A}_{2121}^1\psi_{,2222}- \left(\mathcal{A}_{1111}^1+\mathcal{A}_{2222}^1-2 \mathcal{A}_{1122}^1-2 \mathcal{A}_{1221}^1\right)\psi_{,1122}+\mathcal{A}_{1212}^1 \psi_{,1111}=0.
\end{equation}
We look for a traveling wave solution to \eqref{eq-fourth-order-PDE} in the form  
\begin{equation}\label{sol-peri}
\psi=\zeta\left(n x_2\right) \mathrm{e}^{\mathrm{i} n x_1},
\end{equation}
where $n$ is the wavenumber and `$\mathrm{i}$' is the imaginary unit. Inserting \eqref{sol-peri} into \eqref{eq-fourth-order-PDE} yields a fourth-order ordinary differential equation of $\zeta(nx_2)$ with a  characteristic given by
\begin{equation}
\mathcal{A}_{2121}^1 s^4-\left(\mathcal{A}_{1111}^1+\mathcal{A}_{2222}^1-2 \mathcal{A}_{1122}^1-2 \mathcal{A}_{1221}^1\right) s^2+\mathcal{A}_{1212}^1 =0,
\label{cha-eq}
\end{equation}
where $s$ is the eigenvalue and the general solutions for $\zeta(nx_2)$ and $\hat{\zeta}(nx_2)$ are
\begin{equation}
\begin{aligned}
 &\zeta(n x_2)= \sum_{j=1}^4 Q_j \exp \left(n s_j x_2\right),\hspace{28.5mm}  x_2 \in\left(0, h\right), \\
&\hat{\zeta}(n x_2)=  Q_5 \exp \left(n \hat{s}_1 x_2\right)+Q_6 \exp \left(n \hat{s}_2 x_2\right), \quad x_2 \in(-\infty, 0),
\end{aligned}
\label{sol-general}
\end{equation}
where $Q_1$ to $Q_6$ are constants, and without loss of generality we have removed solutions that do not decay at infinity for the half-space. Hence,  $s_i$ and $\hat{s}_i$ $(i=1,2)$ have  positive real parts. 

From \eqref{inc-bc}, we write down the linearized incremental boundary conditions and continuity conditions: 
\begin{equation}
\left.\begin{aligned}
&T_{i}^{(l)}=0, \quad \text { on } x_{2}=h,\\
&\hat{u}_{i}=u_{i}, \quad \hat{T}_{i}^{(l)}=T_{i}^{(l)}, \quad \text { on } x_{2}=0,
\end{aligned}\right\}\quad i=1,2,
\label{bc-linear}
\end{equation}
where
\begin{equation}
    T_{i}^{(l)}=\mathcal{A}_{2 i l k}^{1} u_{k, l}+\bar{p} u_{2, i}-p^{*} \delta_{2 i},\quad
\hat{T}_{i}^{(l)}=\hat{\mathcal{A}}_{2 i l k}^{1} \hat{u}_{k, l}+\hat{\bar{p}} {\color{black}\hat{u}_{2, i}}-\hat{p}^{*} \delta_{2 i}.
 \label{trantion-linear}
\end{equation}

A direct substitution of \eqref{sol-general} into \eqref{bc-linear} gives rise to the linear system:
\begin{equation}
\mathbf{G}\mathbf{Q}=0,
\end{equation}
where $\mathbf{G}$ is a $6\times6$ matrix whose elements are dependent on the principal stretch $\lambda$ and the material parameters once $W$ and $\hat{W}$ are specified, and $\mathbf{Q}=\left[Q_1,Q_2,Q_3,Q_4,Q_5,Q_6\right]^\mathsf{T}.$ The existence of a non-trivial solution requires
\begin{equation}
\operatorname{det} \mathbf{G}=0,
\label{bif-general}
\end{equation}
which characterizes the values of the bifurcation parameter at which an instability takes place.

We emphasize that the thin film or the half-space is anisotropic. For the isotropic material, we adopt a neo-Hookean law with strain energy function given by
\begin{equation} \label{neo-Hookean}
W_\mathrm{nH} =\frac{\mu_\mathrm{nH}}{2} \left(\operatorname{tr} \mathbf{B}-3\right),
\end{equation}
where $\mu_\mathrm{nH}$ is the shear modulus and $\mathbf{B}$ can be found in $\eqref{Cauchy-Green}_1$. In the subsequent analysis, a subscript `nH' will indicate that the neo-Hookean model \eqref{neo-Hookean} is used.

Anisotropy arises from two families of fibers, as illustrated in Figure \ref{bilayer} for which we adopt the HGO model \citep{Holzapfel2000JE}:
\begin{equation} 
W_\mathrm{HGO}=\frac{\mu_\mathrm{HGO}}{2} \left(I_1-3\right)+\frac{k_1}{2k_2}\bigg(\mathrm{exp}[\left(I_4-1\right)^2 k_2]+\mathrm{exp}[\left(I_6-1\right)^2 k_2]-2\bigg),
\label{HGO-model}
\end{equation}
where $\mu_\mathrm{HGO}$ is also the ground-state shear modulus, $k_1$ corresponds to fiber stiffness (with stress unit), and $k_2$ is a dimensionless parameter that controls the nonlinear response of the fiber. The principal invariants $I_i$ $(i=1,4,6)$ are defined as
\begin{equation}
I_1=\operatorname{tr} \mathbf{C},\quad I_4=\mathbf{M} \cdot \mathbf{C} \mathbf{M},\quad
I_6=\mathbf{M}^{\prime} \cdot \mathbf{C} \mathbf{M}^{\prime},
\label{eq-I4I6}
\end{equation}
where $\mathbf{M}$ and $\mathbf{M}^{\prime}$ are the directions of the two fibers families, and $\mathbf{C}$ is the right Cauchy-Green deformation tensor  $\eqref{Cauchy-Green}_2$. Similarly, the subscript `HGO' is correlated to \eqref{HGO-model}. It can be seen that for $k_1=0$ the HGO model is identical to the neo-Hookean model.

\begin{figure}[!ht]
    \centering
    \includegraphics[width=0.5\textwidth]{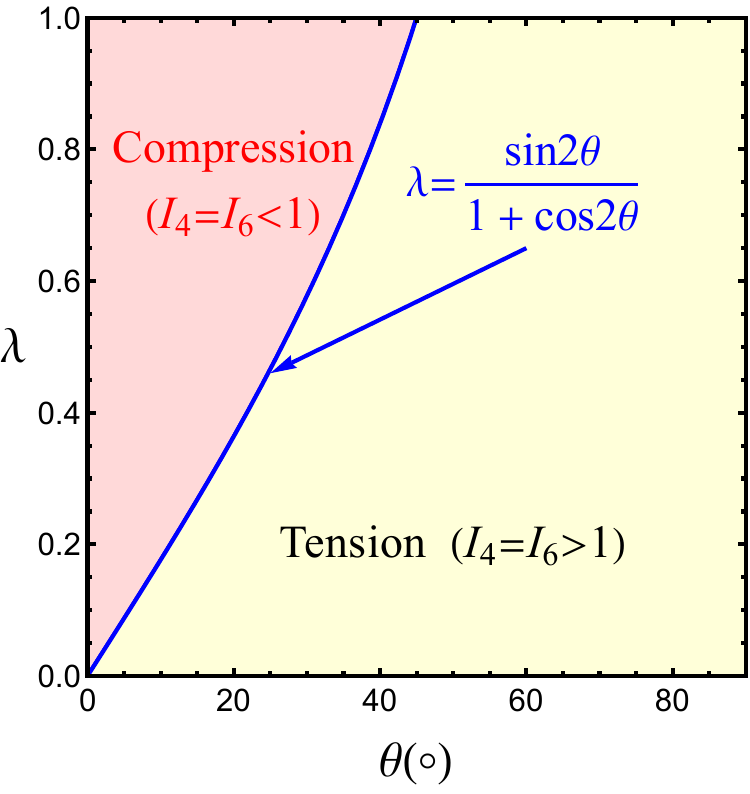}
       \caption{Transition between fiber compression and fiber tension during the uniform deformation.}
       \label{fig:lambda-theta}
 \end{figure}

We assume that the fibers are {\color{black}equal} and opposite with respect  to the $X_1$-axis at an angle $\theta$:
\begin{equation}
\mathbf{M}=\cos \theta \mathbf{e}_1+\sin \theta \mathbf{e}_2, \quad \mathbf{M}^{\prime}=\cos \theta \mathbf{e}_1-\sin \theta \mathbf{e}_2.
\label{eq-vectorM}
\end{equation}
By use of \eqref{def-gradient-hg}, the two principal invariants $I_4$ and $I_6$ in homogeneous deformation are expressed as
\begin{equation}
    I_4=I_6=\lambda^2\cos^2\theta+\dfrac{\sin^2\theta}{\lambda^2}.
\end{equation}
As discussed in \cite{Holzapfel2000JE}, the fiber reinforcement becomes inactive when either $I_4\leqslant1$ or $I_6\leqslant1$, namely, fibers cannot support compressive stresses, in which case the associated response  reduces to the isotropic case. In view of this, we solve the equation $I_4-1=0$ to obtain the transition compression ratio given by
\begin{equation}
    \lambda=\frac{\sin2\theta}{1+\cos2\theta}.
    \label{eq-lambda-theta}
\end{equation}
We further plot in Figure \ref{fig:lambda-theta} the compressive and tension regions for the fibers. The material is reinforced only for fiber angles and compression within the light yellow region. This is the region where we perform our analysis. Otherwise, the neo-Hookean model will be employed and the problem reduces to the standard isotropic wrinkling problem \citep{Alawiye2020JMPS}. Specifically, when $\lambda=1$ the corresponding fiber angle is $\theta=45^\circ$ as derived from \eqref{eq-lambda-theta}.

During a homogeneous deformation, we impose the zero traction condition at $x_2 = h$ and ensure traction continuity at the interface $x_2 = 0$. This condition specifies the Lagrange multipliers $\bar{p}_\mathrm{nH}$ and $\bar{p}_\mathrm{HGO}$ from $\bar{\pi}_{22}=0$ as follows
\begin{equation}
    \begin{aligned}
       & \bar{p}_\mathrm{nH}=\frac{\mu_\mathrm{nH} }{\lambda^2},\\
        & \bar{p}_\mathrm{HGO}=\frac{\mu_\mathrm{HGO}}{\lambda^2} +\frac{4 k_1 \sin^2 \theta}{\lambda^2}\left(\lambda^2 \cos^2 \theta-1+\frac{\sin^2 \theta}{\lambda^2}\right)\left(\mathrm{exp}\left[{k_2\left(\lambda^2 \cos^2 \theta-1+\frac{\sin^2 \theta}{\lambda^2}\right)^2}\right]\right).
    \end{aligned}
    \label{eq-pressure}
\end{equation}
The latter is consistent with the result in \citet{Mirandola2023JAP}. Furthermore, these two expressions are universally valid regardless of the location of neo-Hookean material and the HGO material.
\vskip12pt

\noindent We consider two scenarios:
\begin{itemize}
    \item Case I: a neo-Hookean half-space coated by an HGO film,
    \item Case II: an HGO half-space coated by a neo-Hookean film.
\end{itemize}
To avoid confusion, we emphasize the consistency of notations by making the following connections:
\begin{equation}
   \left\{ \begin{aligned}
        &\text{case I}: \hat{W}=W_\mathrm{nH},~\hat{\bar{p}}=\bar{p}_\mathrm{nH},~\hat{\mu}=\mu_\mathrm{nH},~ W=W_\mathrm{HGO},~\bar{p}=\bar{p}_\mathrm{HGO},~\mu=\mu_\mathrm{HGO},\text{ and so on}, \\
        &\text{case II: }\hat{W}=W_\mathrm{HGO},~\hat{\bar{p}}=\bar{p}_\mathrm{HGO},~\hat{\mu}=\mu_\mathrm{HGO},~ W=W_\mathrm{nH},~\bar{p}=\bar{p}_\mathrm{nH},~\mu=\mu_\mathrm{nH},\text{ and so on}.
    \end{aligned}\right.
    \label{definition}
\end{equation}
\vskip12pt

Further, we introduce the ratio of the shear modulus of the half-space to that of the thin film $r=\hat{\mu}/\mu$. We further measure all stress quantities with respect to $\mu$ , which amounts to take $\mu=1$ and stresses dimensionless without loss of generality. Hence in the subsequent analysis $k_1$ will be dimensionless when it is specified by a value.

When the neo-Hookean material model \eqref{neo-Hookean} is used,  the eigenvalues of \eqref{cha-eq} are given by $s_1=1, s_2=\lambda^2, s_3=-1, s_4=-\lambda^2$. However, for the HGO model \eqref{HGO-model}, the eigenvalues have cumbersome expressions that will not be explicitly presented. Additionally, we may encounter repeated roots that lead to the coefficient matrix $\mathbf{G}$ being multiplied by a zero factor, which will result in extraneous solutions that need proper care.
 
The bifurcation condition \eqref{bif-general} can be recast as
\begin{equation}
    {\color{black}\det\mathbf{G}= \Xi(\lambda,nh,r,k_1,k_2,\theta)=0,}
    \label{bif-exact}
\end{equation}
where $nh$ can be regarded as a unified parameter as they never appear individually \citep{Cai1999PRSA,Alawiye2019PTRSA,Liu2024IJSS}. Now the bifurcation condition \eqref{bif-exact} is related to the principal stretch $\lambda$, the wavenumber $nh$, the modulus ratio $r$, and the anisotropy parameters $(k_1,k_2,\theta)$.

Once the parameters $r$, $k_1$, $k_2$ and $\theta$ are given, it is convenient to plot the bifurcation curve of $\lambda$ versus the wavenumber $nh$. If $r\ll1$, namely, the film is extremely stiffer than the substrate, it is {\color{black}known} that the $\lambda$-$n h$ curve will attain a maximum which identifies the critical stretch $\lambda_\mathrm{cr}$ and the critical wavenumber $(nh)_\mathrm{cr}$ \citep{Cai1999PRSA,Liu2014IJES,Alawiye2019PTRSA}. Furthermore, the critical state can be determined from \eqref{bif-exact} and
\begin{equation}
    \frac{\partial \Xi}{\partial (nh)}=0.
    \label{Dbif}
\end{equation}
As $r$ approaches to 1, i.e., the film and the substrate share a similar modulus and a transition between supercritical and subcritical bifurcations may occur, but this can only be tackled in the nonlinear regime \citep{Cai1999PRSA,Hutchinson2013,Jin2019IJSS,Alawiye2020JMPS}, which will be investigated in the next section.

In the following, we will explore separately the critical state for two scenarios by focusing on the allowable parametric domain $r\ll1$ where surface wrinkling would be more favorable.

\subsection{Case I: HGO film/neo-Hookean substrate}\label{caseI}
We consider the first scenario in which the film is fiber-reinforced. Then, the bifurcation condition \eqref{bif-exact} and equation \eqref{Dbif} will yield the values of the critical stretch $\lambda_\mathrm{cr}$ and the critical wavenumber $(nh)_\mathrm{cr}$ for given material parameters. We find that the parameter $k_2$ has a marginal influence on the critical state so in this subsection it will be given the literature value of $k_2 =0.8393$ \citet{Holzapfel2017}. Further, the effect of the modulus ratio $r$ on the critical state has been studied extensively for film-substrate bilayers \citep{Cai1999PRSA,Hutchinson2013,Cai2019IJNM,Yin2018IJSS,Alawiye2019PTRSA,Nguyen2020BMM,Liu2024IJSS}, so we no longer repeat the analysis for $r$ in the current context and will directly focus on the influence of the parameters related to fibers on the critical state.

\begin{figure}[!ht]
    \centering
    {\subfigure[]{\includegraphics[width=0.45\textwidth]{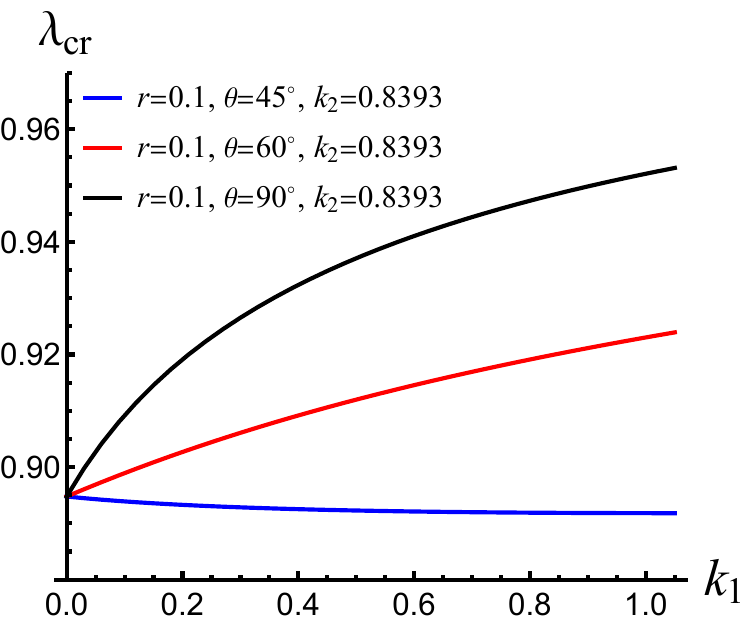}\label{lambdacr-k1}}}
    \hspace{4mm}
    {\subfigure[]{\includegraphics[width=0.45\textwidth]{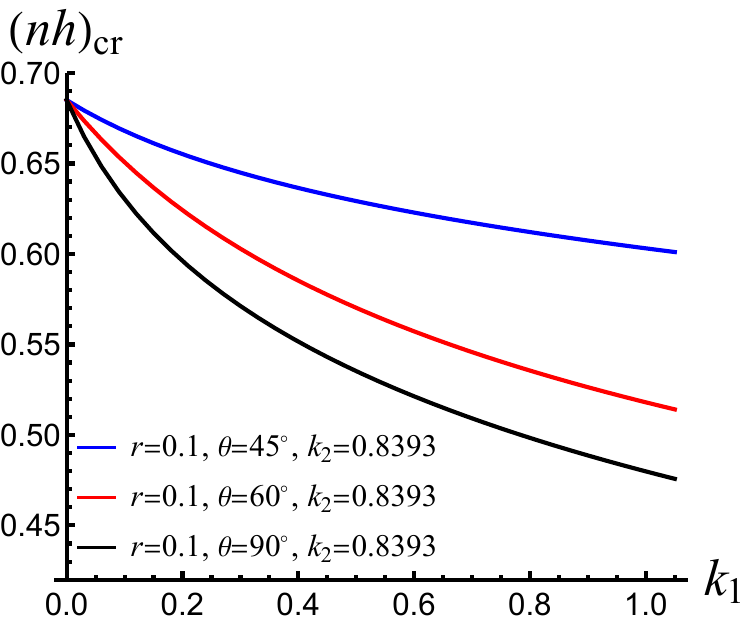}\label{modecr-k1}}}
       \caption{The critical stretch $\lambda_\mathrm{cr}$ and the critical wavenumber $(nh)_\mathrm{cr}$ as functions of $k_1$ when $r=0.1$, $k_2=0.8393$ with different values of $\theta$ for an anisotropic HGO film.
       } 
    \label{cr-k1}
 \end{figure}

Figure \ref{cr-k1} shows the critical stretch $\lambda_\mathrm{cr}$ and the critical wavenumber $(kh)_\mathrm{cr}$ versus fiber stiffness $k_1$. The parameter $r=0.1$ implies that the HGO film is $10$ times stiffer than the neo-Hookean substrate. We select three representative values of the fiber angle, namely $\theta=45^\circ, 60^\circ, 90^\circ$, which implies from Figure \ref{fig:lambda-theta} that no fiber is under compression in the homogeneous deformation. 

As $k_1\rightarrow0$, the bilayer structure will reduce to its isotropic counterpart with $\lambda_\mathrm{cr}=0.8947$ and $(kh)_\mathrm{cr}=0.6849$, regardless of the values of $\theta$ and $k_2$. If $\theta=60^\circ$ or $\theta=90^\circ$,  the critical stretch $\lambda_\mathrm{cr}$ is growing as the fiber stiffness $k_1$ increases as shown in Figure \ref{lambdacr-k1}. Therefore, a stiffer fiber will promote surface wrinkling. 

However, for $\theta=45^\circ$, a reverse trend is observed. Such non-monotonic characteristic in terms of the fiber angle is typical in fiber-reinforced materials, such as in cylindrical tubes \citep{Goriely2013PRSA} and in a film bonded to an anisotropic substrate \citep{Stewart2016EML,Nguyen2020BMM}. In addition, the admissible range of $\lambda_\mathrm{cr}$ for $\theta=90^\circ$ has the maximum amplitude while the counterpart for $\theta=45^\circ$ has the minimum. This implies that the tangent of each curve is also dependent on the fiber angle $\theta$. Figure \ref{modecr-k1} shows the same monotonically decreasing behavior for the critical wavenumber as the fiber stiffness $k_1$ varies. Consequently, the inclusion of fibers creates a pattern with less wrinkles. 

\begin{figure}[!ht]
    \centering
    {\subfigure[]{\includegraphics[width=0.45\textwidth]{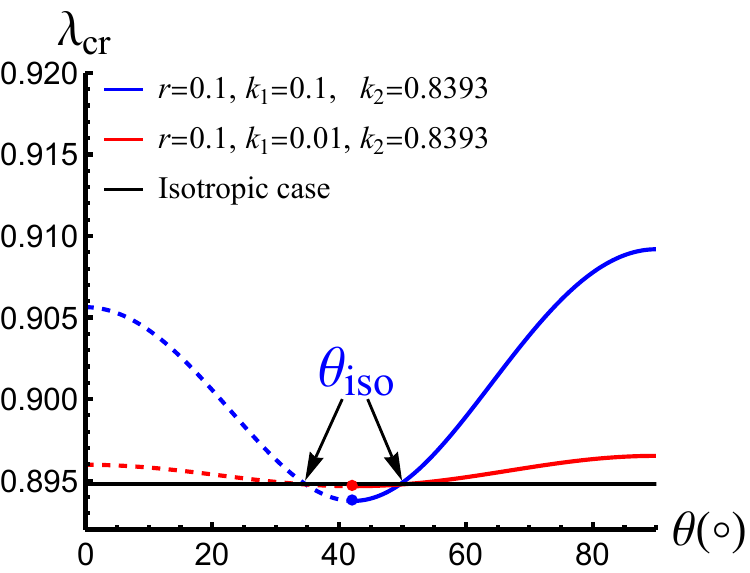}\label{lambdacr-theta}}}
    \hspace{4mm}
    {\subfigure[]{\includegraphics[width=0.45\textwidth]{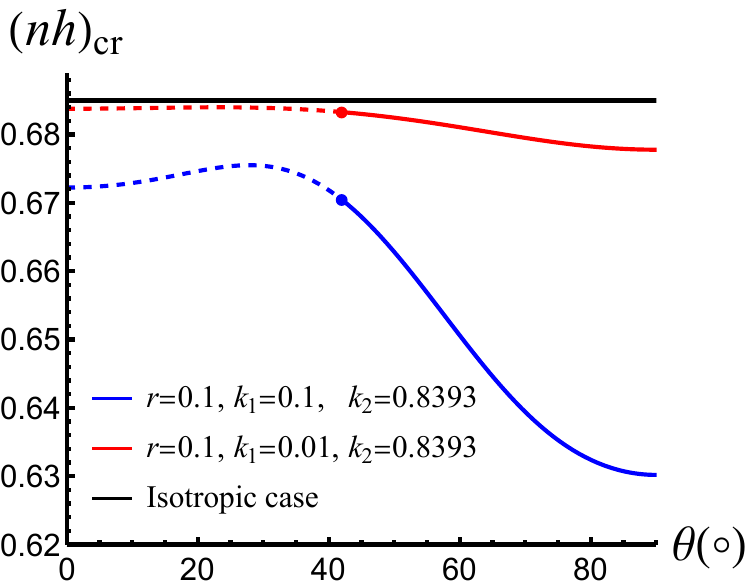}\label{modecr-theta}}}
       \caption{The critical stretch $\lambda_\mathrm{cr}$ and the critical wavenumber $(nh)_\mathrm{cr}$ as functions of fiber angle $\theta$ when $r=0.1$, $k_2=0.8393$ with different values of $k_1$ for an HGO film.}
    \label{cr-theta}
 \end{figure}

We further show the influence of the fiber angle $\theta$   from $0^\circ$ to $90^\circ$  on the critical state in Figure \ref{cr-theta}. The black line corresponds to the isotropic counterpart. It can be seen from Figures \ref{lambdacr-theta} and \ref{modecr-theta} that the curves for $k_1=0.1$ and $0.01$ are similar. In particular, there are two intersection points, marked by $\theta_\mathrm{iso}$, where the critical stretch $\lambda_\mathrm{cr}$ is identical to that of the isotropic structure. We further identify the intersections of the blue and red curves with the transition angle $\theta_\mathrm{tr}$ illustrated in Figure \ref{fig:lambda-theta} and find that $\theta\approx42^\circ$ for both curves, which is also close to the local minimum. The dashed part of each curve indicates that the fibers will be compressive for these angles, and as a result has no physical meaning. Moreover, a stable bilayer is attained if $42^\circ<\theta<\theta_\mathrm{iso}$. In this case, the stiffness of the fiber can be used to delay surface instability. Otherwise, anisotropy promotes the instability, and a stiffer fiber will give rise to an earlier instability. 

In Figure \ref{modecr-theta} we show the monotonic nature of the critical wavenumber $(nh)_\mathrm{cr}$ as a function of $\theta$ in the solid curve. In that case, less wrinkles are created with fibers perpendicular to the compression direction (which is expected physically as wrinkles are then more energy costly to create). A clear feature is that the wavenumber is lower for a higher $k_1$, which is consistent with the results illustrated in Figure \ref{modecr-k1}. 

\begin{figure}[!ht]
    \centering
    {\subfigure[]{\includegraphics[width=0.45\textwidth]{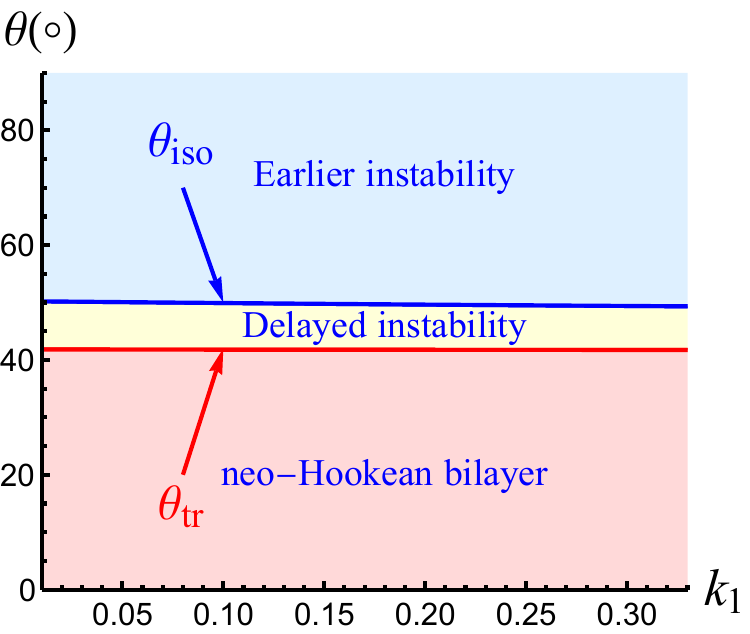}\label{thetaiso-k1}}}
    \hspace{4mm}
    {\subfigure[]{\includegraphics[width=0.45\textwidth]{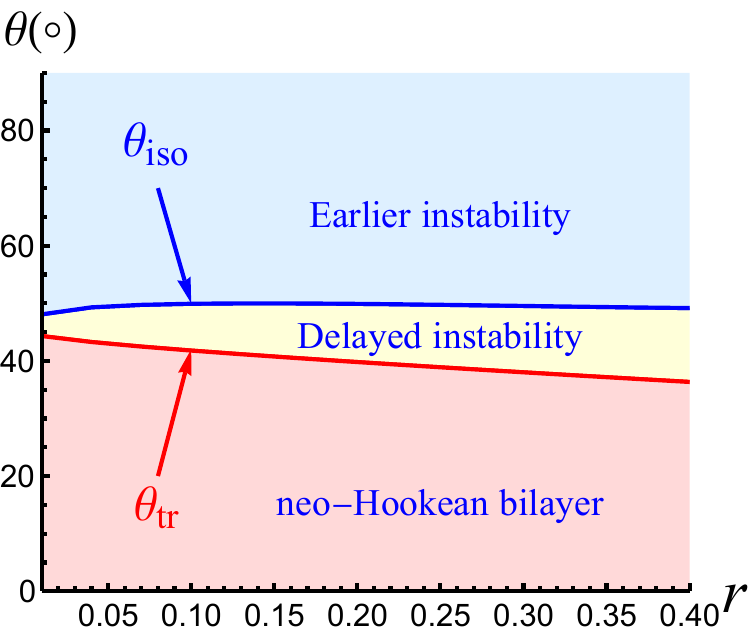}\label{thetaiso-r}}}
       \caption{The transition angles $\theta_\mathrm{iso}$ and $\theta_\mathrm{tr}$ as functions of $k_1$ and $r$. The parameters are given by $k_2=0.8393$, (a) $r=0.1$ and (b) $k_1=0.1$. The film is composed of an HGO material.}
    \label{theta-k1}
 \end{figure}

In {\color{black}general}, the angles $\theta_\mathrm{iso}$ and $\theta_\mathrm{tr}$ are the solutions of 
\begin{equation}
    \begin{aligned}
      &\Xi\left(\lambda_\mathrm{iso},(nh)_\mathrm{iso},r,k_1,k_2,\theta\right)=0, \\& 
     \Xi\left(\lambda,nh,r,k_1,k_2,\theta\right)\Big|_{\lambda=\frac{\sin2\theta}{1+\cos2\theta}}=0,\quad \frac{\partial\Xi\left(\lambda,nh,r,k_1,k_2,\theta\right)}{\partial (nh)}\Big|_{\lambda=\frac{\sin2\theta}{1+\cos2\theta}}=0, 
    \end{aligned}\label{eq-transition}
\end{equation}
where the subscript `iso' indicates the corresponding critical stretch and critical wavenumber for the neo-Hookean bilayer counterpart. The associated results are shown in Figure \ref{theta-k1}. We highlight different regions by colors to indicate the effect of fiber reinforcement on the onset of surface instability. We see that the two transition angles $\theta_\mathrm{iso}$ and $\theta_\mathrm{tr}$ are nearly independent of the fiber stiffness $k_1$. However, the value of $k_1$ will definitely affect the magnitude of the critical stretch away from the isotropic counterpart for a given fiber angle $\theta$, as shown in Figure \ref{lambdacr-theta}. From Figure \ref{thetaiso-r}, we observe that the modulus ratio $r$ has a prominent influence on the lower transition angle $\theta_\mathrm{tr}$. However, the higher transition angle $\theta_\mathrm{iso}$ is almost identical as $r$ changes.

\subsection{Case II: neo-Hookean film/HGO substrate}\label{caseII}
\begin{figure}[!ht]
    \centering
    {\subfigure[]{\includegraphics[width=0.45\textwidth]{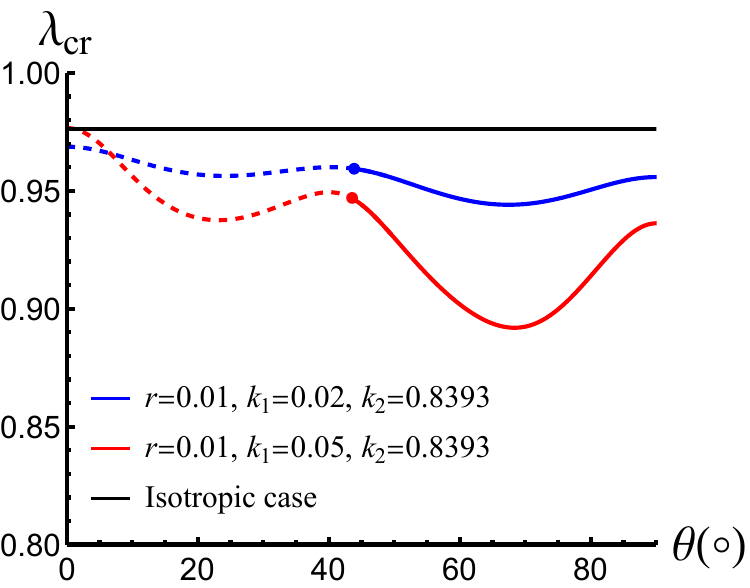}\label{lambdacr-theta-case2}}}
    \hspace{4mm}
    {\subfigure[]{\includegraphics[width=0.45\textwidth]{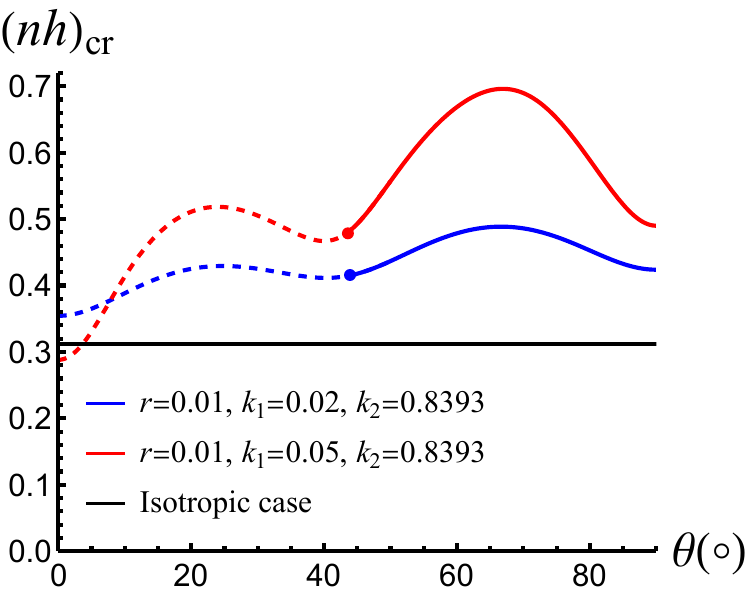}\label{modecr-theta-case2}}}
       \caption{The critical stretch $\lambda_\mathrm{cr}$ and the critical wavenumber $(nh)_\mathrm{cr}$ as functions of fiber angle $\theta$ as $r=0.01$, $k_2=0.8393$ with different values of $k_1$. The substrate is composed of an HGO material. In this case, $r=0.01$ and $k_1=0.02$ result in $\mu_\mathrm{HGO}=0.01\mu_\mathrm{nH}$ and $k_1=0.02\mu_\mathrm{nH}=2\mu_\mathrm{HGO}$.}
    \label{cr-theta-case2}
 \end{figure}

Next, we assume that the film is isotropic while the substrate is fiber-reinforced. Surface wrinkling of such anisotropic bilayers has been extensively studied through both linear analysis and finite element simulations  \citep{Stewart2016EML,Nguyen2020BMM,Guo2023}. Previous studies have uncovered fundamental insights into the effects of fiber properties on the critical state. Hence, we only study the dependence of the critical stretch $\lambda_\mathrm{cr}$ and the critical wavenumber $(nh)_\mathrm{cr}$ on the fiber angle $\theta$ in Figure \ref{cr-theta-case2}.

As shown in Figure \ref{cr-theta-case2}, which recovers the results of \citet{Stewart2016EML} and \citet{Nguyen2020BMM}, the critical stretch and wavenumber are non-monotonic with respect to the fiber angle. Comparisons of Figures \ref{lambdacr-theta-case2} and \ref{modecr-theta-case2} indicate that the $\lambda_\mathrm{cr}$-curve and the $(nh)_\mathrm{cr}$-curve are almost symmetric. For instance, the two local minima of the red curve in Figure \ref{lambdacr-theta-case2} are attained at $\theta\approx23.21^\circ$ and $\theta\approx68.35^\circ$, while the red curve in Figure \ref{modecr-theta-case2} has two local maxima at $\theta\approx24^\circ$ and $\theta\approx66.91^\circ$. The dashed lines indicate the angle values that are excluded in the HGO model as the fibers are in compression. Therefore, we consider values of $\theta$  starting approximately at $44^\circ$. Moreover, if the substrate is fiber-reinforced, the critical stretch curve, as shown in Figure \ref{lambdacr-theta-case2}, is always lower than the black line. Similarly, the associated critical wavenumber is higher than the black line. Thus, fiber reinforcement in the substrate\textit{ delays the instability}  and \textit{results in more wrinkles}.

\begin{figure}[!ht]
    \centering
    \subfigure[]{
    \includegraphics[width=0.45\textwidth]{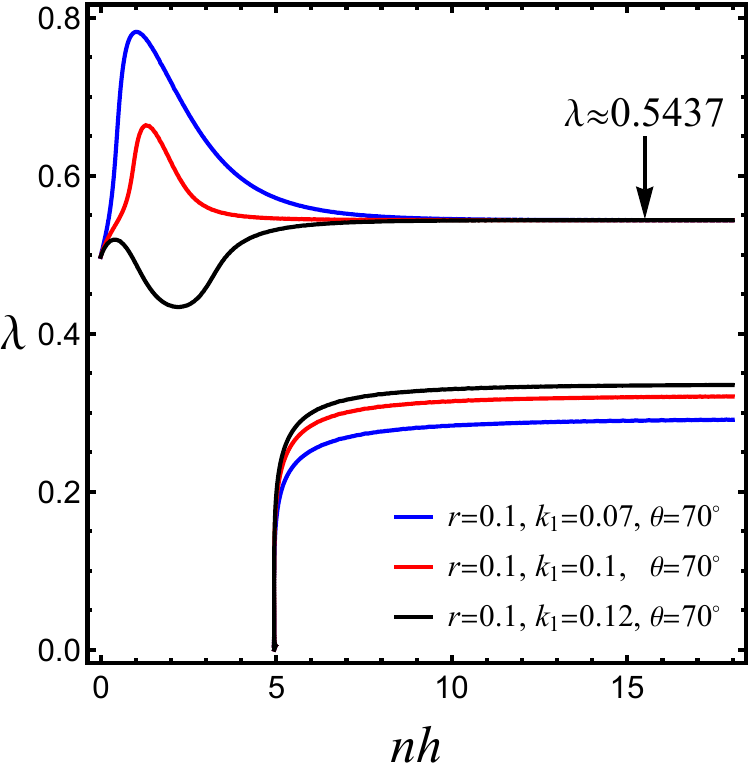}{\label{fig:lambda-nh-k1}}}
    \hspace{4mm}
     \subfigure[]{
    \includegraphics[width=0.45\textwidth]{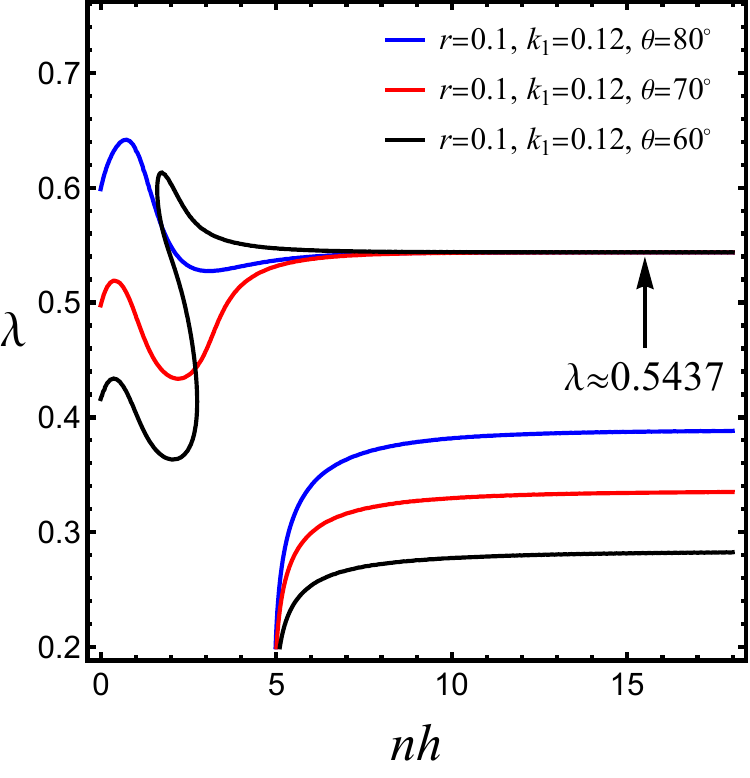}\label{fig:lambda-nh-theta}}
       \caption{Bifurcation curves when fiber stiffness $k_1$ and fiber angle $\theta$ take different values. The parameter $k_2$ is specified by $0$.}
       \label{fig:lambda-nh}
 \end{figure}

In Figure \ref{cr-theta-case2}, we focus on the case where $\mu_\mathrm{nH}$ is much larger than both $k_1$ and $\mu_\mathrm{HGO}$. The fibers still have a considerable influence on the critical state.  In Figure \ref{fig:lambda-nh} we investigate how fiber stiffness and fiber angle affect the bifurcation. We find that the critical stretches for the blue ($k_1=0.07$) and red ($k_1=0.1$) curves are given by $\lambda_\mathrm{cr}\approx0.7821$ and $\lambda_\mathrm{cr}\approx0.6639$, respectively. However, when for larger values of $k_1$  (black curve), the local maximum disappears and instead turns into a local minimum. As a result, the critical stretch is replaced by $\lambda_\mathrm{cr}\approx0.5437$, which is attained as $nh\rightarrow\infty$. This value is the Biot value, originally identified for surface instabilities in a compressed half-space \citep{Biot1963}. We refer to the instability associated with the Biot value and infinite wavenumber as 'Biot instability' to distinguish it from the wrinkling in a bilayer with finite wavenumber. This implies that there exists a special value of $k_1$, across which a mode transition between surface wrinkles and Biot instability takes place. We further display the bifurcations when the fiber angle is varying in Figure \ref{fig:lambda-nh-theta} by taking $k_1=0.12$. The critical stretches for $\theta=80^\circ$ (blue curve) and $\theta=60^\circ$ ({\color{black}black} curve) are found to be $\lambda_\mathrm{cr}\approx0.6417$ and $\lambda_\mathrm{cr}\approx0.6131$, respectively. In particular, the black curve ($\theta=60^\circ$) has two local maxima.
A similar phenomenon was reported when a substrate is coated by a double-layer \citep{Zhou2022MMS}. 

When $\theta=70^\circ$, it is already {\color{black}known} from Figure \ref{fig:lambda-nh-k1} that $\lambda_\mathrm{cr}\approx0.5437$. Similarly, we know that as the fiber angle decreases from $\theta=80^\circ$, the instability mode changes from surface wrinkling (finite wavenumber) to a Biot instability (infinite wavenumber). As the fiber angle further reduces to $\theta=60^\circ$, the Biot instability will give way to surface wrinkling. These facts indicate the existence of two possible fiber angles where transitions between different modes may occur.

Currently, we have shown that there exist several transition fiber angles, i.e., the angle distinguishing an HGO material and a neo-Hookean material, as shown in Figure \ref{fig:lambda-theta}, and the angle distinguishing wrinkles and Biot instability, as shown in Figure \ref{fig:lambda-nh}. The first transition angle can be found in light of $\eqref{eq-transition}_2$. For the latter case, we know that there are two angles. The higher one can be found by 
\begin{equation}
\Xi\left(\lambda,nh,r,k_1,k_2,\theta\right)\Big|_{\lambda=0.5437}=0,\quad \frac{\partial\Xi\left(\lambda,nh,r,k_1,k_2,\theta\right)}{\partial (nh)}\Big|_{\lambda=0.5437}=0,
 \label{eq-transition-wb}
\end{equation}
where $r$, $k_1$, $k_2$ are given and $nh$, $\theta$ are two unknowns. 

\begin{figure}[!ht]
    \centering
    \subfigure[]{
    \includegraphics[width=0.45\textwidth]{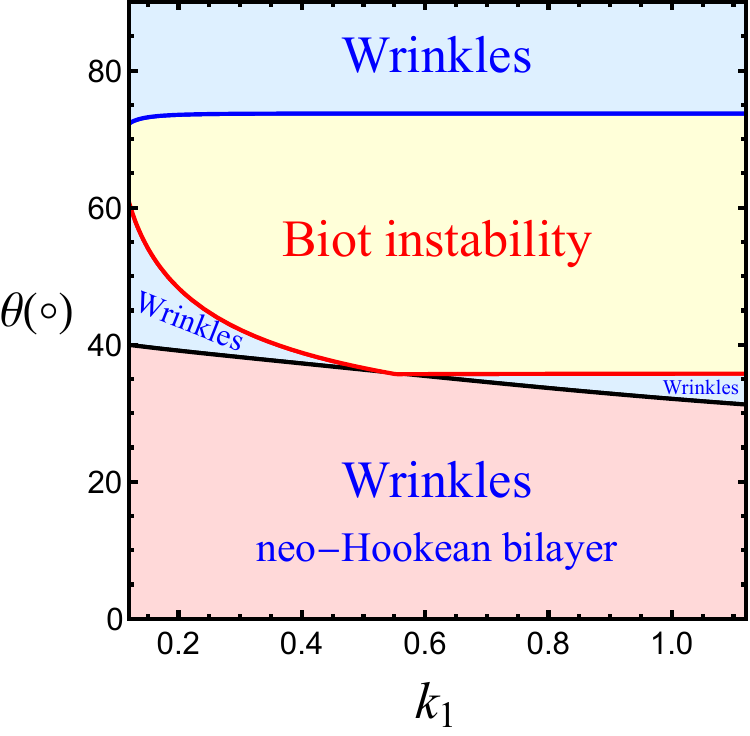}{\label{fig:theta-k1-case2}}}
    \hspace{4mm}
     \subfigure[]{
    \includegraphics[width=0.45\textwidth]{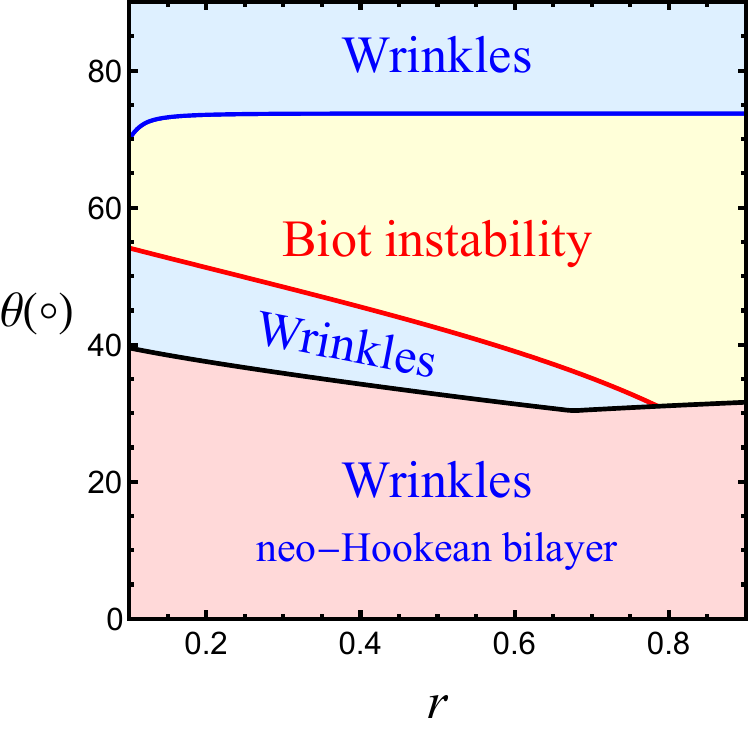}{\label{fig:theta-r-case2}}}
       \caption{Phase diagrams of (a) the fiber angle $\theta$ and fiber stiffness $k_1$ with $r=0.1$, $k_2=0$ and (b) the fiber angle $\theta$ and the modulus ratio $r$ with $k_1=0.15$, $k_2=0$.}
       \label{fig:pha-diag}
 \end{figure}

However, the solution procedure for small angles is hard to identify numerically, because the maximum of black curve in Figure \ref{fig:lambda-nh-theta} coincides with the horizontal line $\lambda\approx0.5437$ at this transition angle. Correspondingly, there is only one critical wavenumber $(nh)_\mathrm{cr}\rightarrow\infty$, and the second equation $\eqref{eq-transition-wb}_2$ fails. To bypass this difficulty, we approximate the solution by moving the stretch from $0.5437$ to $0.5437\times1.01$ (one percent higher) in \eqref{eq-transition-wb}. Finally, we express the phase diagrams in Figure \ref{fig:pha-diag}. The light blue regions correspond to wrinkles of a bilayer where the substrate is fiber-reinforced. The Biot instability is dominant in the light yellow regions, while the bilayer behaves as a neo-Hookean bilayer in the light red domains. We point out that these phase diagrams are merely based on a linear bifurcation analysis and may be subject to modification when the result of nonlinear analysis is taken into consideration.

\section{Weakly nonlinear analysis}\label{nonlinear-analysis}
A linear bifurcation analysis is essential for understanding pattern formation in soft structures. However, insight into post-buckling evolution can only be extracted in the nonlinear regime. 

In this section, we carry out a weakly nonlinear analysis for anisotropic bilayers. Although challenging, a robust theoretical framework in finite elasticity has been established by \cite{Ogden1984}. For homogeneous and isotropic bilayers, post-buckling analyses have been conducted by \cite{Cai1999PRSA,Hutchinson2013,Fu2015PRSA,Cai2019IJNM,Alawiye2020JMPS} for planar structures, and by \cite{Jin2019IJSS,Taffetani2024JMPS} for curved structures.

Similar to the bifurcation analysis, we will elaborate on the solution procedure without distinguishing case I and case II, and only give details for the film where all symbols are presented without hats. 

To perform a nonlinear analysis, we assume the following scaling:
\begin{equation}
\lambda=\lambda_\mathrm{cr}+\varepsilon^{2} \lambda_{1}, \quad \bar{p}=\bar{p}_\mathrm{cr}+\varepsilon^{2} \bar{p}_{1},\label{increment}
\end{equation}
where $\lambda_\mathrm{cr}$ is the critical stretch, $\bar{p}_\mathrm{cr}$ is the pressure of the critical state, $\varepsilon$ is a small parameter characterizing the magnitude of the incremental displacement, and $(\lambda_1,\bar{p}_1)$ are constants of $\mathcal{O}(1)$. In view of $\eqref{increment}_1$, we expand the instantaneous moduli in \eqref{eq:instant moduli} as 
\begin{equation}
    \begin{aligned}
        & \mathcal{A}_{j i l k}^{1}=\left.\mathcal{A}_{j i l k}^{1}\right|_{\lambda=\lambda_\mathrm{cr}}+\varepsilon^2\lambda_1\mathcal{A}_{j i l k}^{1\prime},\\&
        \mathcal{A}_{j i l k n m}^{1}=\left.\mathcal{A}_{j i l k n m}^{1}\right|_{\lambda=\lambda_\mathrm{cr}}+\varepsilon^2\lambda_1\mathcal{A}_{j i l k n m}^{1\prime},\\&
        \mathcal{A}_{j i l k n m q p}^{1}=\left.\mathcal{A}_{j i l k n m q p}^{1}\right|_{\lambda=\lambda_\mathrm{cr}}+\varepsilon^2\lambda_1\mathcal{A}_{j i l k n m q p}^{1\prime},
    \end{aligned}
\end{equation}
where we have defined
\begin{equation}
    \mathcal{A}_{j i l k}^{1\prime}=\left.\frac{\partial\mathcal{A}_{j i l k}^1}{\partial \lambda}\right|_{\lambda=\lambda_\mathrm{cr}},\quad \mathcal{A}_{j i l k n m}^{2\prime}=\left.\frac{\partial\mathcal{A}_{j i l k n m}^2}{\partial \lambda}\right|_{\lambda=\lambda_\mathrm{cr}},\quad \mathcal{A}_{j i l k n m q p}^{3\prime}=\left.\frac{\partial\mathcal{A}_{j i l k n m q p}^3}{\partial \lambda}\right|_{\lambda=\lambda_\mathrm{cr}}.
\end{equation}

In the following, we  keep the original form $\mathcal{A}_{jilk}^{1}$ to denote the associated projection onto $\lambda=\lambda_\mathrm{cr}$ for convenience.

We then look for an asymptotic solution of the form
\begin{equation}
\begin{aligned}
   & u_{i}=\varepsilon u_{i}^{(1)}\left(x_{1}, x_{2}\right)+\varepsilon^{2} u_{i}^{(2)}\left(x_{1}, x_{2}\right)+\varepsilon^{3} u_{i}^{(3)}\left(x_{1}, x_{2}\right)+\cdots, \\
  &p^{*}=\varepsilon p^{(1)}\left(x_{1}, x_{2}\right)+\varepsilon^{2} p^{(2)}\left(x_{1}, x_{2}\right)+\varepsilon^{3} p^{(3)}\left(x_{1}, x_{2}\right)+\cdots.
\end{aligned}\label{sol-expansion}
\end{equation}

Inserting \eqref{sol-expansion} into \eqref{eq-third-order} and \eqref{inc-nonlinear} yields three series equations in $\varepsilon$. The leading-order equations can be obtained from the coefficients of $\varepsilon$ and it can be readily verified that  the resulting system admits the following solutions 
\begin{equation}
u_1^{(1)}=A\zeta^{\prime}(x_2)E+\text{c.c.}, \quad u_2^{(1)}=-\mathrm{i}A\zeta(x_2)E+\text{c.c.},  \quad \quad p^{(1)}=A P_1 E+\text {c.c.},\quad E=\mathrm{e}^{\mathrm{i} x_1},
\label{sol-first}
\end{equation}
where  a prime is defined as a derivative with respect to $x_2$, $A$ is the unknown amplitude, $\zeta$ $(\text{and }\hat{\zeta})$ can be found in \eqref{sol-general}, `$\text{c.c.}$' denotes the complex conjugate of the previous term, and $P_1$ is solved by substituting (\ref{sol-first}) into $(\ref{eq-linear})_1$ (with $i=1$):
\begin{equation}
P_1\left(x_2\right)=-\mathrm{i}\mathcal{A}_{2121}^{1}\zeta^{\prime \prime \prime}+\mathrm{i}\left(\mathcal{A}_{1111}^{1}-\mathcal{A}_{1122}^{1}-\mathcal{A}_{1221}^{1}\right) \zeta^{\prime}.
\end{equation}

We set $n=1$ without loss of generality as $h$ and $n$ always appear in the product $nh$ and we use $h$ to obtain different values of $nh$. In fact, \eqref{sol-first} is the solution of the linear bifurcation analysis with an unknown amplitude written as $A$.

The coefficients of $\varepsilon^2$ provide
\begin{equation}
\begin{aligned}
&u_{1,1}^{(2)}+u_{2,2}^{(2)}=\frac{1}{2} u_{i, j}^{(1)} u_{j, i}^{(1)},\\
&\mathcal{A}_{j i l k}^{1} u_{k, l j}^{(2)}-p_{, i}^{(2)}=-\mathcal{A}_{j i l k n m}^{2} u_{m, n}^{(1)} u_{k, l j}^{(1)}-p_{, j}^{(1)} u_{j, i}^{(1)},\quad i=1,2,\hfill
\end{aligned}\label{eq-second}
\end{equation}
along with the boundary conditions and the continuity conditions 
\begin{equation}
\left.
\begin{aligned}
&T_i^{(2)}=0, \quad \text { on } x_2=h,\\
&u_i^{(2)}=\hat{u}_i^{(2)}, \quad T_i^{(2)}=\hat{T}_i^{(2)},\quad \text {on } x_2=0, \hfill\\
&\hat{u}_i^{(2)} \rightarrow 0, \quad \text { as } x_2 \rightarrow-\infty,
\end{aligned}\right\}\quad i=1,2,
\label{bc-second}
\end{equation}
where
\begin{equation}\label{T_i^{(2)}}
T_i^{(2)}=\mathcal{A}_{2 i l k}^{1\prime} u_{k, l}^{(2)}+\bar{p}_\mathrm{cr} u_{2, i}^{(2)}-p^{(2)} \delta_{2 i}+\frac{1}{2} \mathcal{A}_{2 i l k n m}^{2\prime} u_{k, l}^{(1)} u_{m, n}^{(1)}-\bar{p}_\mathrm{cr} u_{2, k}^{(1)} u_{k, i}^{(1)}+p^{(1)} u_{2, i}^{(1)}.
\end{equation}

In particular, the inhomogeneous terms are related to
\begin{equation}
u_{i, j}^{(1)}=A \Gamma_{i j}^{(1)} E+\text { c.c.},
\label{inh-term-1}
\end{equation}
with 
\begin{equation}
    \Gamma_{11}^{(1)}=\mathrm{i}\zeta^\prime,\quad \Gamma_{12}^{(1)}=\zeta^{\prime\prime},\quad \Gamma_{21}^{(1)}=\zeta,\quad \Gamma_{22}^{(1)}=-\mathrm{i}\zeta^\prime.
\end{equation}
It can be readily checked that the inhomogeneous terms in \eqref{eq-second} are linear combinations of $E^0$, $E^2$, and $E^{-2}$. Therefore, we are able to pursue a particular solution for $(u_1^{(2)},u_2^{(2)},p^{(2)})$ as
\begin{equation}
\begin{aligned}
&u_1^{(2)}=A A^* U_0\left(x_2\right)+A^2 U_2\left(x_2\right) E^2+\text {c.c., } \\
&u_2^{(2)}=A A^*  V_0\left(x_2\right)+A^2 V_2\left(x_2\right) E^2+\text {c.c., }\\
&p^{(2)}=A A^*  P_0+A^2 P_2 E^2+\text {c.c.},
\end{aligned}\label{sol-second}
\end{equation}
where the $*$ denotes the complex conjugate.

To identify the unknown functions $U_0$, $U_2$, $V_0$, $V_2$, $P_0$, $P_2$, we substitute \eqref{sol-second} into \eqref{eq-second}, \eqref{bc-second} and collect the coefficients of $E^0$ and $E^2$. As a result, a well-imposed equation system is obtained. Specifically, many of them can be solved directly to give rise to analytical expressions of $U_0$, $U_2$, $V_0$, $P_0$, and $P_2$, and $V_2$ satisfies the following equation
\begin{equation}
\mathcal{A}_{2121}^{1} V_2^{\prime \prime \prime \prime}-4\left(\mathcal{A}_{1111}^{1}+\mathcal{A}_{2222}^{1}-2 \mathcal{A}_{1122}^{1}-2 \mathcal{A}_{1221}^{1}\right) V_2^{ \prime \prime}+16\mathcal{A}_{1212}^{1}V_2 = \Phi(x_2),\label{eq-v2-general}
\end{equation}
where $\Phi(x_2)$ is related to $\mathcal{A}^{2}_{jilknm}$, $\zeta(x_2)$ and its derivatives, and the long expression is omitted for brevity. After some manipulations we recast \eqref{eq-v2-general} as
\begin{equation}
V_2^{{\prime \prime \prime \prime}}-\left(1+\tau^2\right) \alpha^2 V_2^{{\prime \prime}}+\tau^2 \alpha^4 V_2=\phi(x_2).
\label{eq-V2}
\end{equation}
Note that $\phi=\Phi/\mathcal{A}_{2121}^{1}$ and we have denoted the square of eigenvalues of the homogeneous part of \eqref{eq-v2-general} by $\alpha^2$ and $\alpha^2\tau^2$, respectively.

A particular solution of \eqref{eq-V2} can be obtained by using Laplace transforms  \citep{Fu2015SIAM}:
\begin{align}
V_\mathrm{PS}&=\frac{1}{2 \alpha^3 \notag \tau\left(1-\tau^2\right)}\left\{\tau \mathrm{e}^{\alpha x_2} \int_0^{x_2} \mathrm{e}^{-\alpha t} \phi(t) \mathrm{d}  t-\mathrm{e}^{\alpha \tau x_2} \int_0^{x_2} \mathrm{e}^{-\alpha \tau t} \phi(t) \mathrm{d}  t\right. \\
&\left.+~\mathrm{e}^{-\alpha \tau x_2} \int_{-\infty}^{x_2} \mathrm{e}^{\alpha \tau t} \phi(t) \mathrm{d}  t-\tau \mathrm{e}^{-\alpha x_2} \int_{-\infty}^{x_2} \mathrm{e}^{\alpha t} \phi(t) \mathrm{d}  t\right\}.
\label{V_ps}
\end{align}
Laplace transforms are not applicable to the layer. Alternatively, the method of variation of parameters can be applied. It is found that by simply changing the lower limit from $\infty$ to $0$ in the particular integral of \eqref{V_ps}, the same expression as that derived from the variation of parameters method can be obtained. Ultimately, the general solution of $V_2$ can be written as 
\begin{equation}
V_2\left(x_2\right)=d_1 \mathrm{e}^{\alpha x_2}+d_2 \mathrm{e}^{\alpha \tau x_2}+d_3 \mathrm{e}^{-\alpha x_2}+d_4 \mathrm{e}^{-\alpha \tau x_2}+V_\mathrm{PS},
\label{sol-V2}
\end{equation}
where $d_i$ $(i=1,2,3,4)$ are constants to be determined from the boundary conditions and $\alpha$, $\alpha\tau$ take the positive real parts. For the half-space, $d_3$ and $d_4$ must vanish according to the decaying condition at infinity.

With the use of $\eqref{sol-second}_{1,2}$, we obtain
\begin{equation}
u_{i, j}^{(2)}=AA^\textrm{c} \Gamma_{i j}^{(m)}+A^2 \Gamma_{i j}^{(2)} E^2+\text {c.c.},
\label{inh-term-2}
\end{equation}
where 
\begin{equation}
\begin{aligned}
& \Gamma_{11}^{(m)}=0, \quad \Gamma_{12}^{(m)}=U_0^{\prime}, \quad \Gamma_{21}^{(m)}=0, \quad \Gamma_{22}^{(m)}=V_0^{\prime}, \\
& \Gamma_{11}^{(2)}=2 \mathrm{i} U_2, \quad \Gamma_{12}^{(2)}=U_2^{\prime}, \quad \Gamma_{21}^{(2)}=2 \mathrm{i} V_2^{\prime}, \quad \Gamma_{22}^{(2)}=V_2^{\prime}.
\end{aligned}
\end{equation}

For the neo-Hookean model, we are able to write explicitly the solutions as 
\begin{equation}
\begin{aligned}
&U_0=0, \quad V_0=2 \zeta \zeta^{\prime}, \quad P_0=2 \mu_\mathrm{nH} \lambda_\mathrm{cr}^{-2}\left(\zeta \zeta^{\prime}\right)^{\prime}+2 \mathrm{i} P_1 \zeta^{\prime}, \\
&2 \mathrm{i} U_2=-V_2^{\prime}+\zeta \zeta^{\prime \prime}-\zeta^{\prime 2}, \quad 2 \mathrm{i} P_2=\mu_\mathrm{nH} \lambda_\mathrm{cr}^{-2} U_2^{\prime \prime}-4 \mu_\mathrm{nH} \lambda_\mathrm{cr}^2 U_2+P_1^{\prime} \zeta-P_1 \zeta^{\prime}, \\
&V_\textrm{PS}=\frac{\lambda_\mathrm{cr}^4-1}{9\lambda^8-82\lambda^4+9}\left(9\zeta\zeta^{\prime\prime\prime}-21\zeta^\prime\zeta^{\prime\prime}\right).
\end{aligned}
\end{equation}
However, for the HGO model, the counterparts are lengthy and hence are omitted here.

At this point, we are in a position to determine the value unknown amplitude $A$. To achieve this, we refer to the methodology proposed by \cite{Fu1995JE} and \cite{Fu1996QJMAM} and employ the principle of virtual work
\begin{equation}
\int_{-\infty}^h\int_0^{2\pi} \chi_{i j} u_{i,j}^\circ \mathrm{d}x_1\mathrm{d}x_2=0,
\label{virtual-work}
\end{equation}
where $u_i^\circ$ corresponds to the linear solutions with $n=-1$ and precisely possesses the form of
\begin{equation}
\begin{aligned}
&u_1^\circ=\zeta^{\prime}\left(x_2\right) \mathrm{e}^{-\mathrm{i} x_1}, \quad u_2^\circ=\mathrm{i} \zeta\left(x_2\right) \mathrm{e}^{-\mathrm{i} x_1}, \quad0\leqslant x_2 \leqslant h,\\
&\hat{u}_1^\circ=\hat{\zeta}^{\prime}\left(x_2\right) \mathrm{e}^{-\mathrm{i} x_1}, \quad \hat{u}_2^\circ=\mathrm{i} \hat{\zeta}\left(x_2\right) \mathrm{e}^{-\mathrm{i} x_1}, \quad-\infty<x_2 \leqslant 0.
\end{aligned}
\end{equation}
It is again emphasized that when an integration is carried out within the region occupied by the substrate a hat will be added on the corresponding quantity. We, however, disregard hats in illustrating the derivation of amplitude equation for simplicity, as seen in \eqref{virtual-work}.

On substituting \eqref{sol-expansion} into \eqref{exp-chi} and then {\color{black}inserting} the resulting expression into \eqref{virtual-work}, we can collect the coefficients of $\varepsilon$,  $\varepsilon^2$, and $\varepsilon^3$. It is found that the first two equations by equating the coefficients of $\varepsilon$ and $\varepsilon^2$ to zeros are automatically true. The coefficient of $\varepsilon^3$ leads to 
\begin{equation}
\int_{-\infty}^h\int_0^{2\pi}\left\{\left(\mathcal{A}_{jilk}^1u_{k,l}^{(3)}+\bar{p}_\mathrm{cr}u_{j,i}^{(3)}-p^{(3)}\delta_{ji}+\bar{p}_1 u_{j, i}^{(1)}+  \lambda_1 \mathcal{A}_{j i l k}^{1\prime} u_{k, l}^{(1)}+\sigma_{ij}^{(3)}\right)u_{i,j}^\circ \right\}\mathrm{d}x_1\mathrm{d}x_2=0,
\label{epsilon-3}
\end{equation}
where
\begin{equation}
\sigma_{i j}^{(3)}=\mathcal{A}_{j i l k n m}^2 u_{k, l}^{(1)} u_{m, n}^{(2)}+\frac{1}{6} \mathcal{A}_{j i l k n m q p}^3 u_{k, l}^{(1)} u_{m, n}^{(1)} u_{p, q}^{(1)}+p^{(1)} u_{j, i}^{(2)}+p^{(2)} u_{j, i}^{(1)}.
\end{equation}

Our next step is to eliminate $u_i^{(3)}$ and $p^{(3)}$ in \eqref{epsilon-3}. For this purpose, we repeatedly use integration by parts for terms containing $u_i^{(3)}$ and $p^{(3)}$ based on the fact that $u_i^\circ$ is a linear solution and by means of the identity obtained from the coefficients of $\varepsilon^3$ of \eqref{inc-nonlinear}, namely,
\begin{equation}
    u_{i,i}^{(3)}=u_{i,j}^{(1)}u_{j,i}^{(2)}.
\end{equation}
After some manipulations, we arrive at
\begin{equation}
\int_{-\infty}^h  \int_0^{2 \pi} \left\{\bar{p}_1 u_{j, i}^{(1)} u_{i, j}^\circ+  \lambda_1 \mathcal{A}_{j i l k}^{1\prime} u_{k, l}^{(1)} u_{i, j}^\circ+p^\circ u_{i, j}^{(1)} u_{j, i}^{(2)}+\sigma_{i j}^{(3)} u_{i, j}^\circ\right\} \mathrm{d} x_1\mathrm{d} x_2=0.
\label{integral-final}
\end{equation}
where the pressure field $p^\circ$ is associated with $u_i^\circ$ and is given by
\begin{equation}
p^\circ=\left[\mathrm{i}\mathcal{A}_{2121}^{1}\zeta^{\prime \prime \prime}-\mathrm{i}\left(\mathcal{A}_{1111}^{1}-\mathcal{A}_{1122}^{1}-\mathcal{A}_{1221}^{1}\right) \zeta^{\prime}\right]\mathrm{e}^{-\mathrm{i}x_1}.
\end{equation}

Substituting $\eqref{sol-first}_3$, \eqref{inh-term-1}, $\eqref{sol-second}_3$ and \eqref{inh-term-2} into \eqref{integral-final}, making use of the fact that $u_{i,j}^\circ=\Gamma_{ij}^{(1)*}E^{-1}$, and computing the integral over $x_1$, we deduce the amplitude equation
\begin{equation}
    c_{01}\lambda_1A+c_{03}|A|^2A=0,
    \label{eq-amplitude}
\end{equation}
where $c_{01}$ and $c_{03}$ take the following forms:
\begin{equation}
    \begin{aligned}
        &c_{01}=\int_{-\infty}^h\left\{\mathcal{A}^{1\prime}_{jilk}\Gamma_{kl}^{(1)}\Gamma_{ji}^{(1)*}+\frac{p_1}{\lambda_1}\Gamma_{ij}^{(1)}\Gamma_{ji}^{(1)*}\right\}\mathrm{d}x_2,\\
        &c_{03}=\int_{-\infty}^h\left\{P^*_1\left(\Gamma_{i j}^{(1)} \Gamma_{j i}^{(m)}+2 \Gamma_{i j}^{(1)*} \Gamma_{j i}^{(2)}\right)+K_{i j} \Gamma_{i j}^{(1)*}\right\} \mathrm{d} x_2,
    \end{aligned}\label{eq-c0c1}
\end{equation}
with
\begin{equation}
 K_{i j}=\mathcal{A}_{j i l k n m}^{2}\left(\Gamma_{k l}^{(1)} \Gamma_{m n}^{(m)}+\Gamma_{k l}^{(1)*} \Gamma_{m n}^{(2)}\right)+\frac{1}{2} \mathcal{A}_{j i l k n m q p}^{3} \Gamma_{k l}^{(1)*} \Gamma_{m n}^{(1)} \Gamma_{p q}^{(1)}+P_1 \Gamma_{j i}^{(m)}+P_0 \Gamma_{j i}^{(1)}+P_2\Gamma_{ji}^{(1)*}.
\end{equation}

Equation \eqref{eq-amplitude} admits a non-trivial solution:
\begin{equation}
|A|^2=-\frac{\lambda_1}{c_1},\quad \mbox{with }c_1\equiv\frac{c_{03}}{c_{01}},
\label{am-eq}
\end{equation}
which determines the nonlinear evolution of the amplitude $A$ for a given mode $nh$.

The amplitude equation presented above has a rational solution if and only if the ratio $\lambda_1/c_1$ is negative. This condition establishes the criterion to determine the bifurcation nature: the bifurcation is \textit{supercritical} when $c_1$ is positive and \textit{subcritical} otherwise \citep{Hutchinson1970}. In the subsequent analysis, we consider the two situations separately to study the transition between supercritical and subcritical bifurcations based on \eqref{eq-c0c1}.

Bearing in mind that the parameter $k_2$ has a minor impact on the critical state according to the linear bifurcation analysis, we fix $k_2=0$ in the subsequent analysis. As a result, the reduced strain energy form of \eqref{HGO-model} reads
\begin{equation}
W_\mathrm{HGO}=\frac{\mu_\mathrm{HGO}}{2} \left(I_1-3\right)+\frac{k_1}{2}\bigg(\left(I_4-1\right)^2 +\left(I_6-1\right)^2\bigg).
\label{HGO-model-new}
\end{equation}

\subsection{Case I: HGO film/neo-Hookean substrate}\label{caseI-nonlinear}
\begin{figure}[!ht]
    \centering
    {\subfigure[]{\includegraphics[scale=0.48]{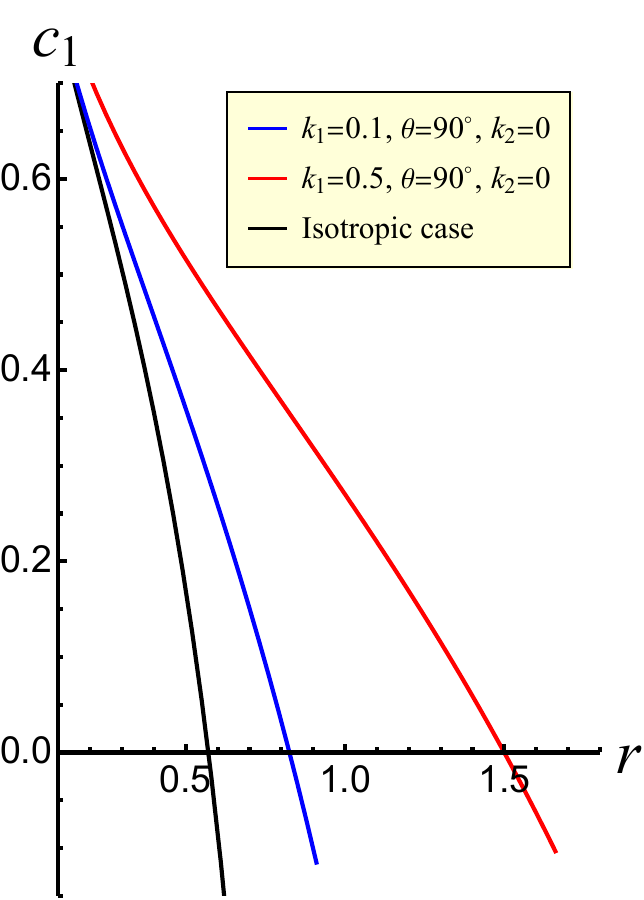}\label{fig:c1-r-case1}}}
    \hspace{2mm}
   {\subfigure[]{\includegraphics[scale=0.48]{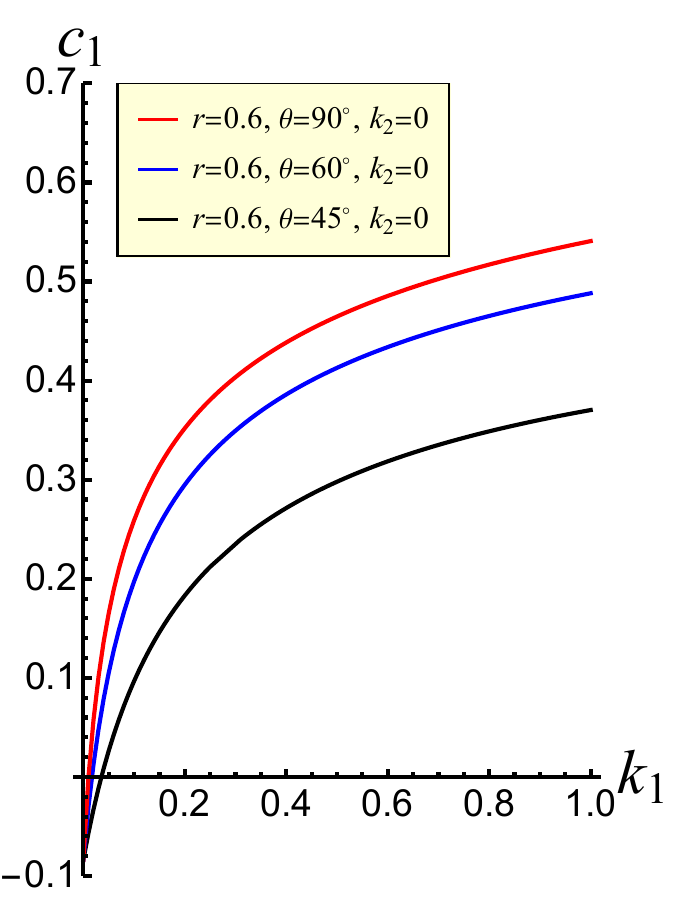}\label{fig:c1-k1-case1}}}
      \hspace{2mm}
     {\subfigure[]{\includegraphics[scale=0.48]{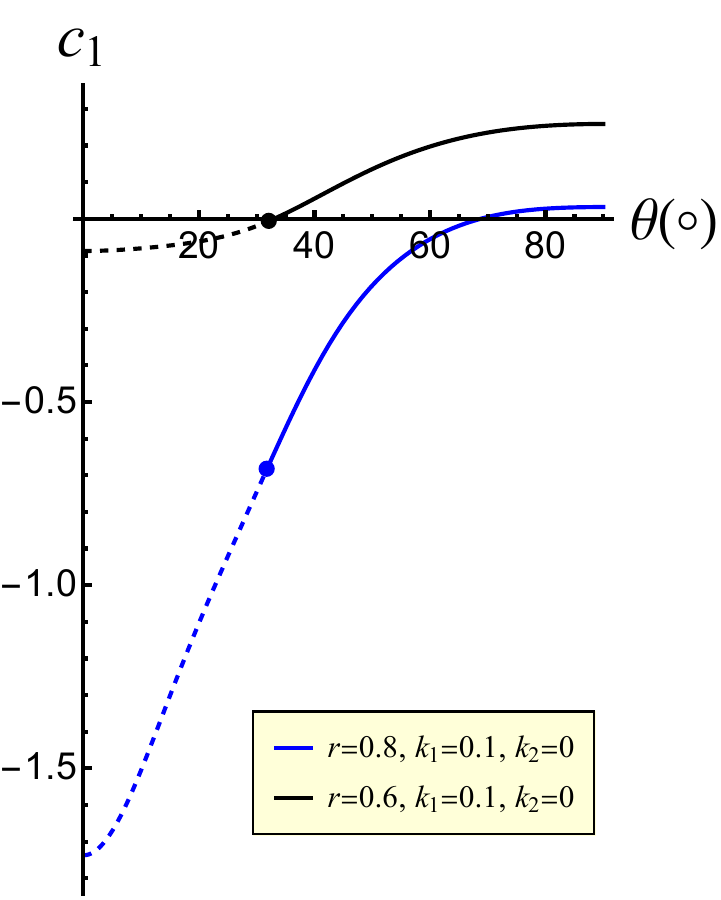}\label{fig:c1-theta-case1}}}
       \caption{Dependence of the coefficient $c_1$ on (a) the modulus ratio $r$, (b) the fiber stiffness $k_1$, and (c) the fiber angle $\theta$. The film is fiber-reinforced and the substrate is isotropic. All used parameters are depicted in each figure.}
    \label{fig:c1-case1}
 \end{figure}

For an anisotropic film, In Figure \ref{fig:c1-case1} we show the coefficient $c_1$ against various parameters. For comparison, the isotropic structure is also shown (black curve). In each curve, the zero corresponds to the transition between supercritical bifurcation ($c_1>0$) and subcritical bifurcation ($c_1<0$). For the isotropic case, this transition happens when $r\approx0.571$ \citep{Alawiye2020JMPS}. As $k_1$ increases, this critical transition value increase. For instance, the zeros for $k_1=0.1$ is at $r\approx0.825$ (blue curve) and at $r\approx1.499$ for $k_1=0.5$ (red curve). We infer that increased fiber stiffness will result in a larger transition modulus ratio.

To further confirm the above conclusion, we show the results of $c_1$ as $k_1$ varies in Figure \ref{fig:c1-k1-case1} where $r=0.6$ and $\theta$ takes the values $45^\circ$, $60^\circ$, $90^\circ$. It can be seen that all curves start from the same point $c_1=-0.0836$ when $k_1=0$, corresponding to the isotropic case, and increase monotonically. Therefore, for $r=0.6>0.571$, initially the bifurcation is subcritical and with increased fiber stiffness the subcritical bifurcation is replaced by a supercritical bifurcation. The monotonic function in $k_1$ implies that $c_1$ reaches a minimum at $k_1=0$, regardless of the fiber angle $\theta$. We conclude that $r\approx0.571$ is a lower bound of the transition value of $r$ and a subcritical bifurcation is impossible for $r<0.571$.

{\color{black}Figure \ref{fig:c1-theta-case1} shows the variation of the coefficient $c_1$ as a function of the fiber angle $\theta$. The dashed curves represent the regime in which fibers are under compression in a homogeneous deformation and are thus excluded from our analysis. The blue and black points correspond to $\theta\approx32^\circ$. Interestingly, each solid curve increases monotonically with respect to $\theta$. Originally, $c_1$ is negative for both cases, resulting in a subcritical bifurcation. As $\theta$ increases, there is a transition from a subcritical  to a supercritical bifurcation.

We define $r_\mathrm{tr}$ as the critical value of $r$   at which a transition from supercritical bifurcation to subcritical occurs, and plot the dependence of $r_\mathrm{tr}$ on the fiber stiffness $k_1$ for $\theta=45^\circ$, $60^\circ$, and $90^\circ$ in Figure \ref{fig:rtr-k1-case1}. As $k_1\rightarrow0$, all curves intersect at $r_\mathrm{th}\approx0.571$. When the fiber orientation is fixed, a subcritical bifurcation, which is sensitive to any imperfection, can occur when the modulus ratio $r$ is below $r_\mathrm{tr}$. Otherwise, a supercritical bifurcation will be triggered at a critical load. We find that the black curve ($\theta=45^\circ$) is always lower than other two curves. The red ($\theta=90^\circ$) curve is initially a little bit higher than the blue one ($\theta=60^\circ$) until $k_1=0.303941$. After that, the blue curve becomes higher.


\begin{figure}[!ht]
    \centering
    \includegraphics[width=4in]{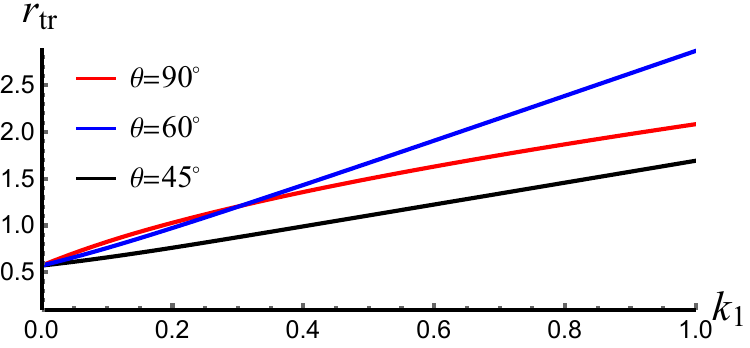}
       \caption{{\color{black}Dependence of the transition value $r_\mathrm{tr}$ on the fiber stiffness $k_1$ with different fiber angles. The film is fiber-reinforced while the substrate is isotropic.
       }}
       \label{fig:rtr-k1-case1}
 \end{figure}

 As expected, all curves are monotonically increasing as a function of $k_1$ (see Figure \ref{fig:rtr-k1-case1}). This is  due to the fact that the film is fiber-reinforced and ab increase in the fiber stiffness is  equivalent to stiffening the film. For this reason, the single modulus ratio $r=\mu_\mathrm{nH}/\mu_\mathrm{HGO}$ loses its physical relevance. Naturally, both the fiber angle $\theta$ and the fiber stiffness will be involved in an effective modulus ratio $\bar{r}$. To identify this, we look at the stress-strain solution in the homogeneous deformation and calculate the limit  when $\lambda\rightarrow1$  to  obtain the  effective shear modulus: 
\begin{equation}
\mu_\mathrm{eff}=\mu _{\mathrm{HGO}}+k_1\left(1+\cos4\theta\right).
\label{eq-eff-mod}
\end{equation} 
Since the reduced strain energy function \eqref{HGO-model-new} is adopted, the parameter $k_2$ does not appear in the above expression. Note that in this formulation, $k_1$ retains its original unit of stress. For $\theta = 90^\circ$, we find $\mu_\mathrm{eff} = \mu_{\mathrm{HGO}} + 2k_1$, whereas for $\theta = 45^\circ$, we have $\mu_\mathrm{eff} = \mu_{\mathrm{HGO}}$. It follows from \eqref{HGO-model-new} that, for fixed $\mu_{\mathrm{HGO}}$ and $k_1$, increasing $\theta$ from $45^\circ$ to $90^\circ$ leads to a larger value of $1 + \cos 4\theta$, thereby stiffening the film. This, in turn, results in an earlier onset of wrinkling instability with fewer wrinkles, which qualitatively explains the results shown in Figure \ref{cr-theta}. We therefore define an effective modulus ratio as
\begin{equation}
\bar{r}=\frac{\mu_\mathrm{nH}}{\mu_\mathrm{eff}}=\frac{r}{1+k_1\left(1+\cos4\theta\right)}.
\label{eq-eff-mod-90}
\end{equation}

By using this new ratio, we plot again Figure \ref{fig:rtr-k1-case1}  in Figure \ref{fig:reff-k1-case1}. It can be seen that the red curve $(\theta=90^\circ)$ approaches $0.571$ with an almost uniform correction as $k_1$ increases. This confirms that the effective modulus \eqref{eq-eff-mod} performs well in this special fiber orientation. However, \eqref{eq-eff-mod} is not adequate to fully capture the combined effects of $\theta$ and $k_1$, as the curves associated with $\theta = 45^\circ$ and $\theta = 90^\circ$ still show noticeable deviations from $0.571$. A more detailed discussion on this point will be provided in Section \ref{discussions}.}

\begin{figure}[!ht]
    \centering
    \includegraphics[width=4in]{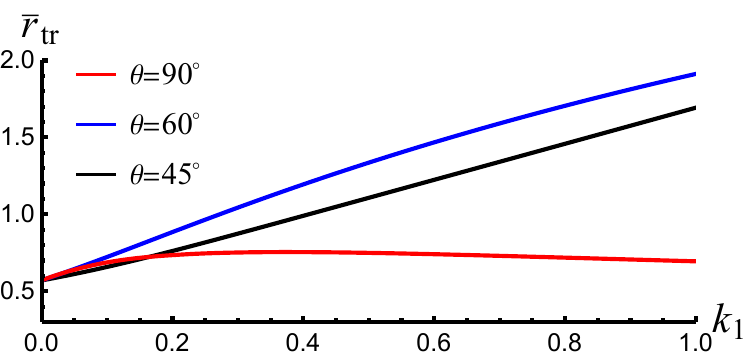}
       \caption{{\color{black}Dependence of the modified transition value $\bar{r}_\mathrm{tr}$ on the fiber stiffness $k_1$ with different fiber angles. The film is fiber-reinforced while the substrate is isotropic.}
       }
       \label{fig:reff-k1-case1}
 \end{figure}

We emphasize that the near-critical evolution of surface wrinkles can be determined from our analytical solution. Obviously, the post-buckling deformation is also involved with the deformation of fibers. Therefore, either compression or tension may be possible for fibers. To investigate this, we refer to \eqref{deformation-relation}, \eqref{eq-I4I6}, and \eqref{eq-vectorM} to write
\begin{align}
\nonumber I_4 &=\frac{\lambda_\mathrm{cr}^2}{2}\left(\left(u_{1,1}+1\right)^2+u_{2,1}^2\right)(1+\cos2\theta)+\frac{1}{2\lambda_\mathrm{cr}^2}\left(\left(u_{2,2}+1\right)^2+u_{1,2}^2\right)(1-\cos2\theta)\\&+
\left(u_{1,2}+u_{2,1}+u_{1,1}u_{1,2}+u_{2,1}u_{2,2}\right)\sin2\theta,\label{eq-I4-post}\\
\nonumber I_6 &=\frac{\lambda_\mathrm{cr}^2}{2}\left(\left(u_{1,1}+1\right)^2+u_{2,1}^2\right)(1+\cos2\theta)+\frac{1}{2\lambda_\mathrm{cr}^2}\left(\left(u_{2,2}+1\right)^2+u_{1,2}^2\right)(1-\cos2\theta)\\&-
\left(u_{1,2}+u_{2,1}+u_{1,1}u_{1,2}+u_{2,1}u_{2,2}\right)\sin2\theta,
\label{eq-I6-post}
\end{align}
where we have defined $u_{1,1}=\partial u_1/\partial x_1$, $u_{1,2}=\partial u_1/\partial x_2$, etc. Furthermore, we substitute the displacement $u_i$ ($i=1,2$) by the leading-order counterparts $\varepsilon u_i^{(1)}$ and absorb the small parameter $\varepsilon$ into the amplitude $A$ according to
\begin{equation}
    \varepsilon |A|=\sqrt{\frac{\lambda_\mathrm{cr}-\lambda}{c_1}}.
    \label{eq:amplidue-sqrt}
\end{equation}
Consequently, we are able to characterize $I_4$ and $I_6$ in the post-buckling regime. 

When the fiber angle is $\theta=90^\circ$, it can be deduced from \eqref{eq-I4-post} and \eqref{eq-I6-post} that $I_4=I_6=\left(\left(u_{2,2}+1\right)^2+u_{1,2}^2\right)/\lambda_\mathrm{cr}^2$. Taking $r=0.01$ and $k_1=0.5$, we determine $\lambda_\mathrm{cr}=0.9852$, $(nh)_\mathrm{cr}=0.2476$ and $c_1=0.9595$. Note that in our post-buckling analysis the wavenumber has been chosen by $n=1$ without loss of generality. So actually the minimum period of the post-buckling solution is $2\pi$. For a given compressive stretch $\lambda<\lambda_\mathrm{cr}$, we define the function 
\begin{equation}
    \mathcal{I}(x_1,x_2)=I_4-1=\frac{\left(u^{(1)}_{2,2}+1\right)^2+\left(u^{(1)}_{1,2}\right)^2}{\lambda_\mathrm{cr}^2}-1,
    \label{eq-I4I6-1st}
\end{equation}
and find that it attains the same minimum at $(0,h)$ or $(2\pi,h)$ within the period $[0,2\pi]\times[0,h]$. Additionally, we ascertain that once $\lambda_\mathrm{cr}-\lambda=0.00365$, the fibers begin to experience compression. Therefore, we define an admissible domain where post-buckling evolution using HGO model is valid. We further consider $r=0.1$ and $k_1=0.1$. In this scenario, we identify $\lambda_\mathrm{cr}=0.9091$, $(nh)_\mathrm{cr}=0.6347$ and $c_1=0.7723$. By using a similar procedure, we find that the condition where fibers start to undergo compression is given by $\lambda_\mathrm{cr}-\lambda=0.01962$. 

\begin{figure}[!ht]
    \centering
    \includegraphics[width=4in]{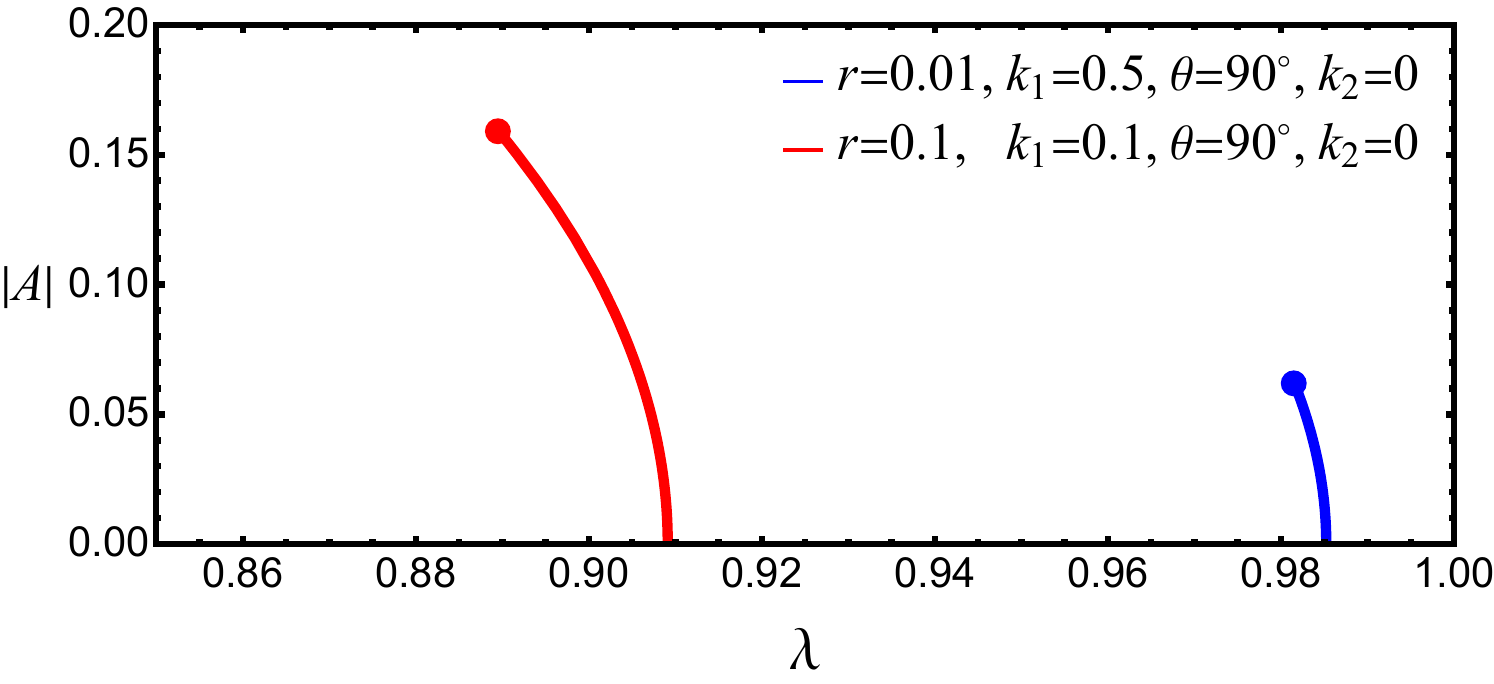}
       \caption{Bifurcation diagrams based on the post-buckling solution derived in this section when the film is fiber-reinforced. Pitchfork supercritical bifurcations are found for both scenarios.}
    \label{fig:A-lambda-case1}
 \end{figure}

Finally, we present the associated bifurcation diagrams for these two cases in Figure \ref{fig:A-lambda-case1}. The blue curve corresponds to $r=0.01$ and $k_1=0.5$ while the red curve represents $r=0.1$ and $k_1=0.1$. Compressive loading begins at $\lambda=1$ and initially the structure undergoes homogeneous deformation, resulting in a trivial amplitude. When compression reaches the critical value $\lambda_\mathrm{cr}$, where the curve intersects the horizontal axis, bifurcation occurs and triggers surface wrinkling. As the structure is further compressed, the amplitude increases, following the path shown in the figure. We also highlight the endpoints where fibers begin to experience compression, marking the limits where the current post-buckling solution becomes invalid. We conclude that the fibers experience compression earlier if they are stiffer.

\subsection{Case II: neo-Hookean film/HGO substrate}\label{caseII-nonlinear}

\begin{figure}[!ht]
    \centering
    {\subfigure[]{\includegraphics[scale=0.475]{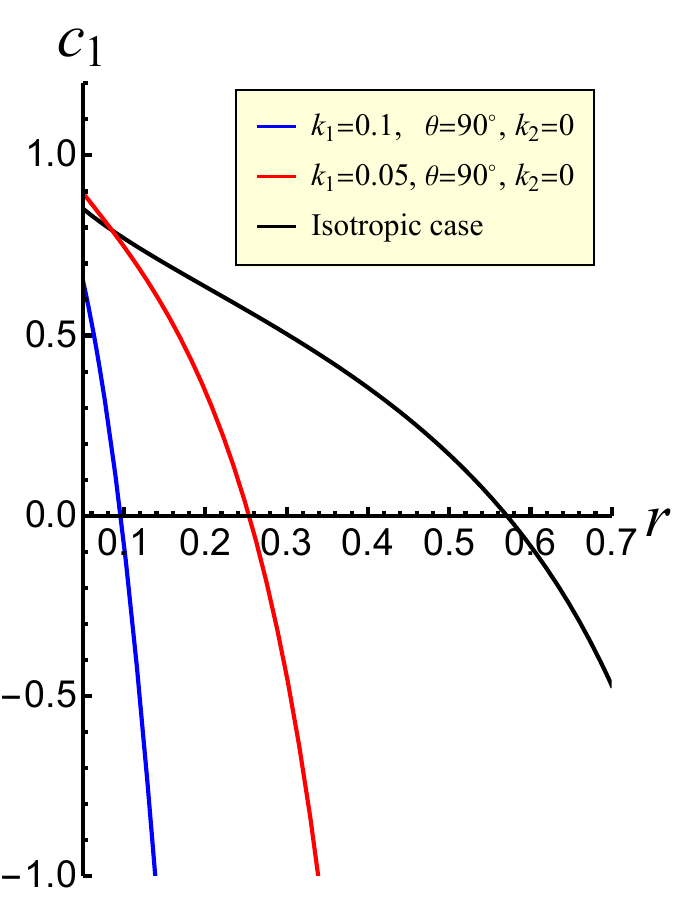}\label{fig:c1-r-case2}}}
    \hspace{2mm}
   {\subfigure[]{\includegraphics[scale=0.475]{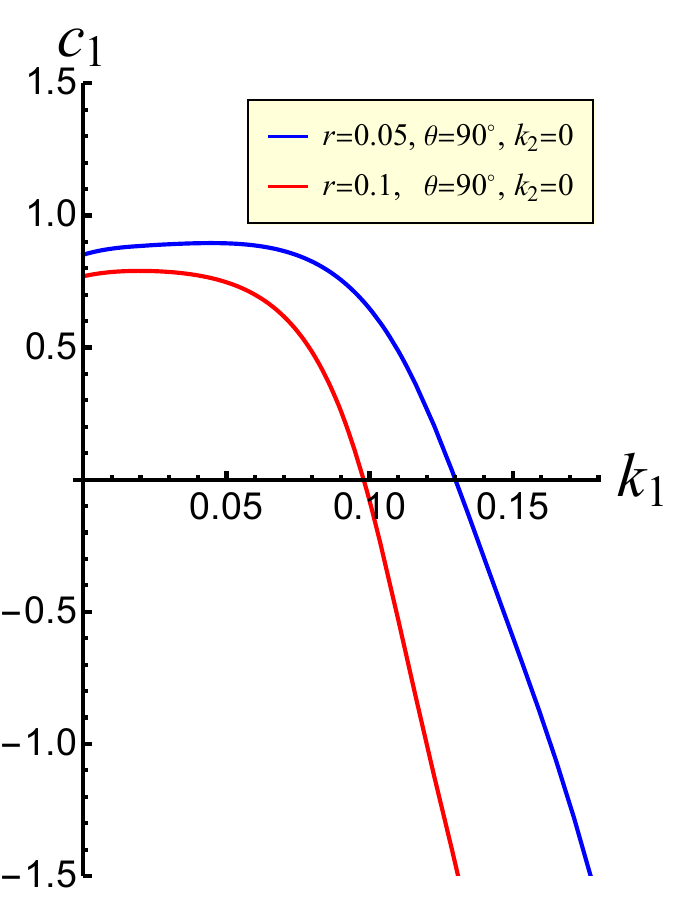}\label{fig:c1-k1-case2}}}
       \hspace{2mm} {\subfigure[]{\includegraphics[scale=0.475]{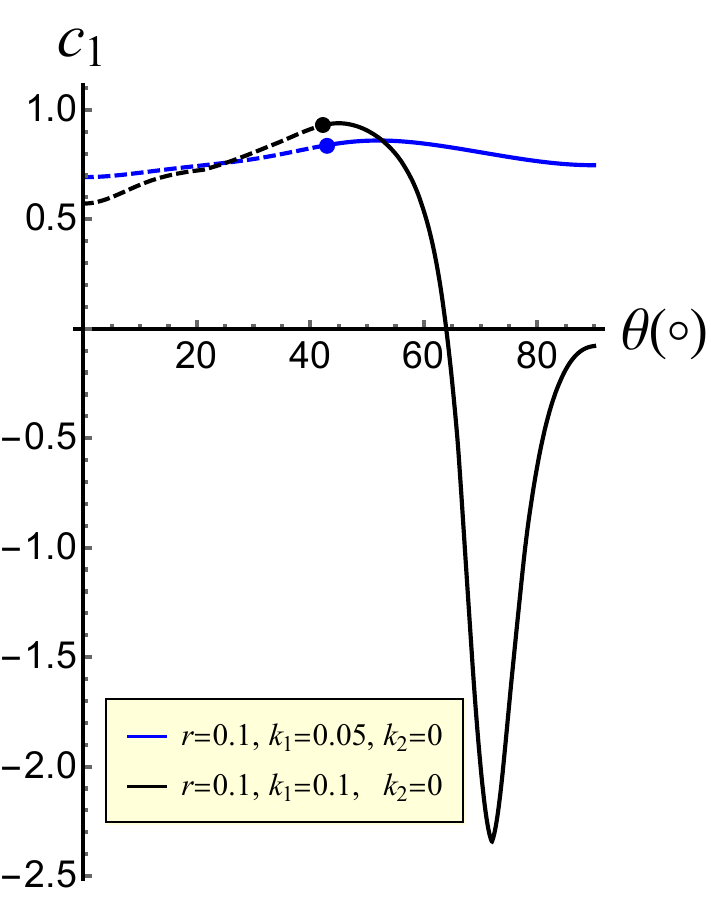}\label{fig:c1-theta-case2}}}
       \caption{Dependence of the coefficient $c_1$ on (a) the modulus ratio $r$, (b) the fiber stiffness $k_1$, and (c) the fiber angle $\theta$. The film is isotropic and the substrate is fiber-reinforced. All used parameters are depicted in each figure.}
    \label{fig:c1-case2}
 \end{figure}

For an anisotropic substrate, we use the same post-buckling solution and plot the coefficient $c_1$ against multiple parameters in Figure \ref{fig:c1-case2}. As before, the intersections with the horizontal axis represent the transition between supercritical and subcritical bifurcations. For $k=0$ (black curve), which corresponds to an isotropic bilayer, we already know that the zero is $r\approx0.571$. The zeros for $k_1=0.05$ and $k_1=0.1$ (red curve) are found to be $r\approx0.254$ and $r\approx0.096$, respectively. Therefore,  a stiffer fiber corresponds to a lower transition modulus ratio, exhibiting a behavior that contrasts with the previous case. When $r\sim0.1$, the black curve intersects with the red one, indicating the non-monotonic characteristic of $c_1$ as a function of $k_1$. 

To clarify the influence of $k_1$, we fix $r$ and $\theta=90^\circ$ in Figure \ref{fig:c1-k1-case2}. Since the modulus ratios differ for the two curves, they start at different values of $c_1$ when $k_1=0$. As expected, both curves slowly rise and then suddenly drop. For $r=0.05$ (blue curve), the supercritical bifurcation gives way to the subcritical one when $k_1\approx0.13$. However, the transition value for $r=0.1$ (red curve) occurs at $k_1\approx0.098$. This manifests that stiffer fibers are needed to induce subcritical {\color{black}bifurcation} if the neo-Hookean film is also stiffer. 

Figure \ref{fig:c1-theta-case2} depicts the relation of $c_1$ and the fiber angle $\theta$. The dashed curve represents fibers under compression and is therefore excluded. If $k_1=0.05$, the effective angle starts at $\theta\approx43^\circ$ and the dependence of $c_1$ on $\theta$ is non-monotonic with $\theta=90^\circ$ corresponding to a local minimum. Meanwhile, the bifurcation is always supercritical. Nonetheless, when the fiber increases to $k_1=0.1$, the effective angle starts from $\theta\approx42^\circ$ and the coefficient $c_1$ varies more significantly with regard to $\theta$. Notably, there exists a transition angle at $\theta\approx64^\circ$ where the bifurcation shifts from supercritical to subcritical. Additionally, the curve reaches a minimum at $\theta\approx72^\circ$ and the tangent around this point is quite sharp. 

From the above analysis, we conclude that for $r>0.571$ a subcritical bifurcation takes place when the substrate is fiber-reinforced and representative transition modulus ratios are given in Table \ref{table}. As expected, with increasing fiber stiffness, the bilayer experiences a subcritical bifurcation at a lower modulus ratio.  Specifically, when $k_1=0.18$ the bifurcation is subcritical even for $r=0.015$. In this case, we know $\mu_\mathrm{nH}=66.7\mu_\mathrm{HGO}$ and $k_1=12\mu_\mathrm{HGO}$ so it can be seen that the fiber stiffness is dominant in the strain energy function \eqref{HGO-model-new}. Therefore, the shear modulus ratio is not an adequate representation of the stiffness mismatch between the two layers.

\begin{figure}[!ht]
    \centering
    \includegraphics[width=4in]{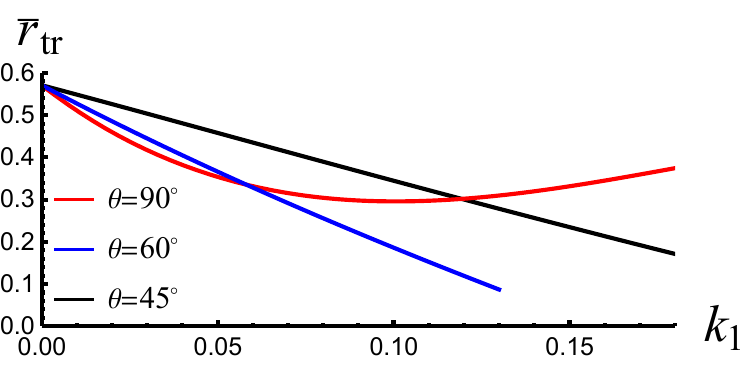}
       \caption{{\color{black}Dependence of the modified transition value $\bar{r}_\mathrm{tr}$ on the fiber stiffness $k_1$ with different fiber angles. The substrate is fiber-reinforced while the film is isotropic.}
       }
       \label{fig:reff-k1-case2}
 \end{figure}

{\color{black}To study the combined contribution of  the shear modulus of the matrix and the fiber stiffness to the effective substrate stiffness, we use the effective shear modulus  \eqref{eq-eff-mod} and define the effective modulus ratio as:
\begin{equation}
\bar{r}=\frac{\mu_\mathrm{eff}}{\mu_\mathrm{nH}}=r+k_1\left(1+\cos4\theta\right).
\label{eq-eff-ratio-case2}
\end{equation}
We show the modified transition value $\bar{r}_\mathrm{tr}$ versus the fiber stiffness $k_1$ for $\theta = 90^\circ$, $60^\circ$, and $45^\circ$ in Figure \ref{fig:reff-k1-case2}. We see that the red curve ($\theta = 90^\circ$) is non-monotonic and deviates significantly from the typical value $r = 0.571$. In contrast, the black ($\theta = 45^\circ$) and blue ($\theta = 60^\circ$) curves exhibit monotonic decrease. Specifically, the blue curve terminates at around $k_1 = 0.13$, beyond which surface wrinkling gives way to a Biot instability, as shown in Figure \ref{fig:theta-k1-case2}. In summary, a supercritical bifurcation arises when $\bar{r} < \bar{r}_\mathrm{tr}$, beyond which it transitions to a subcritical bifurcation. Moreover, the intricate structure of these curves underscores the significant effect of fiber reinforcement.} 

\begin{table}
\begin{center}
\begin{tabular}{|c|c|c|c|c|c|}
    \hline
     $k_1$ & 0.04 & 0.07 & 0.1 & 0.18\\
    \hline
    $r_\mathrm{tr}$ & 0.3020 & 0.1753 & 0.0957 &0.0144\\
    \hline
\end{tabular}
\end{center}
   \caption{Some values of $r_\mathrm{tr}$ where transition between supercritical and subcritical bifurcations takes place for given $k_1$ and $\theta=90^\circ$.}
    \label{table}
\end{table}

\begin{figure}[!ht]
    \centering
    \includegraphics[width=4in]{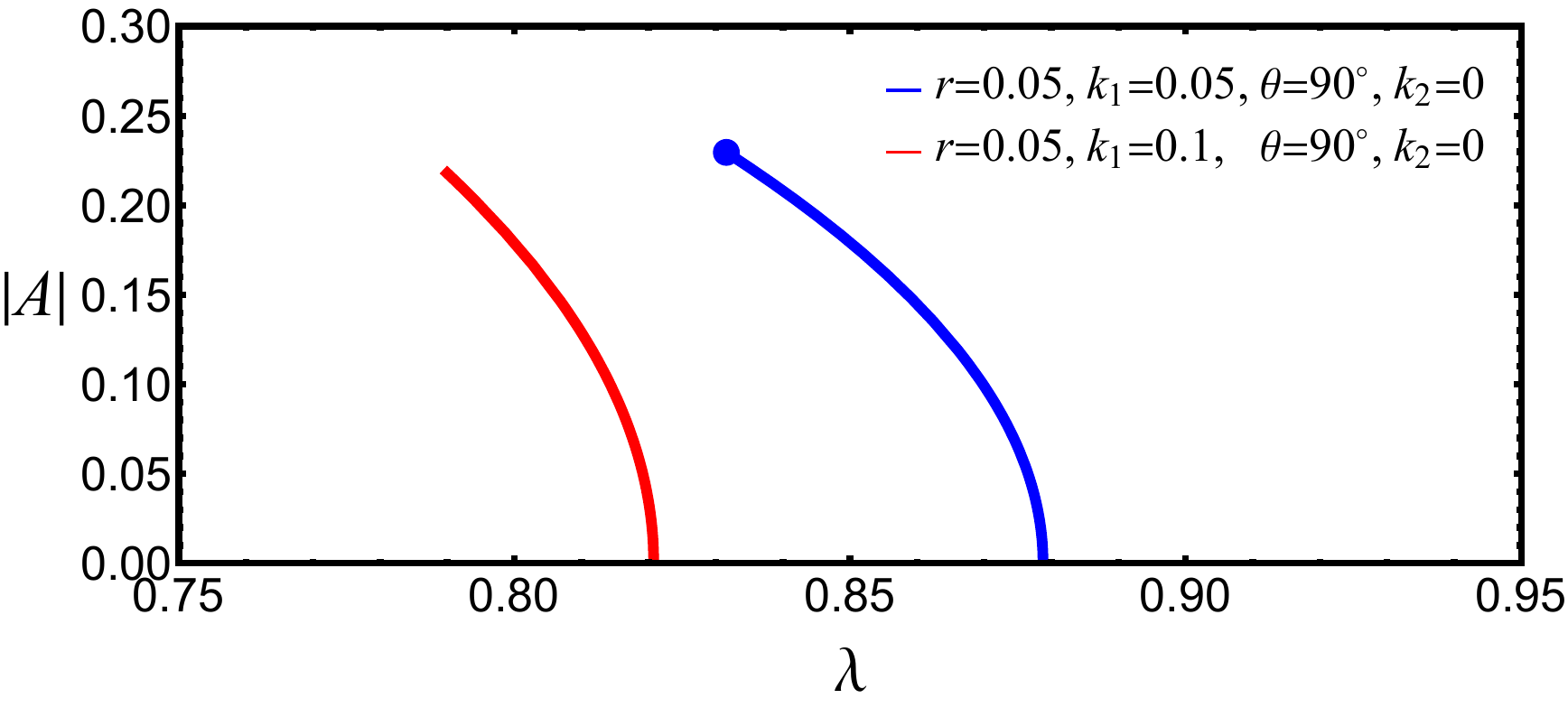}
       \caption{Bifurcation diagrams based on the post-buckling solution derived in this section when the substrate is fiber-reinforced. Pitchfork supercritical bifurcations are found for both scenarios.}
    \label{fig:A-lambda-case}
 \end{figure}

As before, we can identify a domain where our post-buckling solution is applicable when the substrate is fiber-reinforced. We prescribe $r=0.05$ and consider two cases, i.e., either $k_1=0.05$ or $k_1=0.1$. In both cases, the bifurcations are supercritical. We find $\lambda_\mathrm{cr}=0.8788$, $(nh)_\mathrm{cr}=0.6996$ and $c_1=0.8945$ for $r=0.05$ and $k_1=0.01$. Further, the function $\mathcal{I}(x_1,x_2)$ reaches a minimum at $(\pi,0)$ within the period $[0,2\pi]\times(-\infty,0]$. This gives $\lambda_\mathrm{cr}-\lambda=0.0472$, indicating the point where the fibers first experience compression. For $r=0.05$ and $k_1=0.1$, we solve $\lambda_\mathrm{cr}=0.8208$, $(nh)_\mathrm{cr}=0.7625$ and $c_1=0.6477$. Meanwhile, the condition where the fibers are under compression reads $\lambda_\mathrm{cr}-\lambda=0.1333$. Based on these results, we show the associated bifurcation diagrams in Figure \ref{fig:A-lambda-case}. The blue curve for $k_1=0.05$ stops at $\lambda=0.8316$ where the amplitude $|A|\approx0.23$. However, the red curve remains valid in the sense that no fiber is under compression in our analysis until $|A|\approx 0.454$, which is far from the threshold. Therefore, the solution based on the weakly nonlinear analysis may lose validity before reaching the end point. For this reason, we only show the red curve up to $\lambda=0.79$. It can be observed that when the substrate is fiber-reinforced, stiffer fibers experience compression much later.

{\color{black}
\section{Discussions}\label{discussions}
In the preceding sections, we  derived analytically both the bifurcation condition and the amplitude equation governing post-buckling behavior, based on the nonlinear incremental theory of finite elasticity \citep{Ogden1984, Goriely2017}. Notably, the amplitude equations \eqref{eq-amplitude} and \eqref{am-eq} are obtained without introducing any \textit{ad hoc} assumptions or fitting parameters. In this section, we compare our results with  the literature and discuss potential extensions of our work.
\subsection{Comparisons with existing results}
We apply the virtual work method to derive the amplitude equation at third-order nonlinearity \citep{Fu1996QJMAM}. The same result can alternatively be obtained using the Lyapunov–Schmidt–Koiter (LSK) method, originally introduced by \citet{Koiter1945} and subsequently developed by \citet{Thompson1963, Thompson1973}. A detailed exposition of the LSK method is also provided in \citet{Potier-Ferry1987, Triantafyllidis1992ijss}. Post-buckling solutions obtained from both approaches exhibit good agreement with finite element simulations across various case studies \citep{Jin2019IJSS, chen2020continous}. In particular, for a neo-Hookean film bonded to a neo-Hookean substrate under uniaxial compression, \citet{Cai1999PRSA} and \citet{Hutchinson2013} employed the virtual work and LSK methods, respectively, to derive the amplitude equation, both identifying a transition from supercritical to subcritical bifurcations at approximately $r\approx0.571$. In the two scenarios considered in our study, the coefficient $c_1$ depends on the parameters $k_1$ and $\theta$ that characterize the fiber properties. Setting $k_1=0$ reduces the HGO model to the neo-Hookean model, thereby recovering the transition value associated with an isotropic neo-Hookean bilayer, as illustrated in Figures \ref{fig:c1-r-case1} and \ref{fig:c1-r-case2}. This agreement provides a first validation of our analytical post-buckling solutions.

Several previous studies on surface wrinkling of fiber-reinforced bilayers based on the theory of finite elasticity are particularly relevant to the present work \citep{Nguyen2020BMM, Altun2023MoSM, Mirandola2023JAP}. In \citet{Altun2023MoSM}, the film is reinforced with fibers, whereas in \citet{Nguyen2020BMM} and \citet{Mirandola2023JAP}, reinforcement is applied to the substrate. Although the compression considered in \citet{Altun2023MoSM} is growth-induced and the reinforcement involves a single family of fibers aligned in the in-plane direction, qualitative comparisons can still be drawn with our first case. Specifically, \citet{Altun2023MoSM} studied fibers oriented at $\theta = 90^\circ$, aligned along the $X_3$-direction rather than the $X_2$-direction as shown in Figure \ref{bilayer}. Our buckling analysis in Section \ref{caseI} demonstrates that, for a fixed fiber orientation $\theta = 90^\circ$, increasing fiber stiffness destabilizes the structure, reducing the number of wrinkles and thereby increasing the wrinkle wavelength, consistent with the trends observed in Figure 7 of \citet{Altun2023MoSM}. Furthermore, \citet{Altun2023MoSM} developed a numerical scheme to trace the post-buckling evolution and pattern transitions; however, determining the nature of the bifurcation remains challenging without analytical guidance. In practice, random geometric perturbations are often introduced to initiate bifurcation at the critical load. We emphasize that the analytical post-buckling solution \eqref{eq:amplidue-sqrt} not only clarifies the role of fiber stiffness in the transition between subcritical and supercritical bifurcations but also provides an effective initial guess for numerical simulations. As reported by \citet{Dai2013primary}, finite element schemes can converge within five iterations when initialized with an analytical solution.

When the substrate is anisotropic, a detailed linear analysis can be found in  \citet{Nguyen2020BMM,Mirandola2023JAP}, that focuses on the case where the thin film is much stiffer than the substrate. In particular, \citet{Mirandola2023JAP} derived leading-order expressions for the critical stretch and critical wavelength in the limit of a very stiff film. Accordingly, we present only one representative case study in Section \ref{caseII}, and all resulting bifurcation curves show qualitative agreement with the findings in \citet{Nguyen2020BMM,Mirandola2023JAP}. Nevertheless, we extend the discussion by revealing additional bifurcation phenomena in Figure \ref{fig:pha-diag} for the case where the film is only moderately stiffer. These results provide insights into how various parameters influence mode transitions and can serve as a design guideline for tailoring buckling patterns in specific structures.

The main feature of the post-buckling evolution is that, close to bifurcation, the amplitude grows proportionally to the square root of the incremental strain, as shown in equation \eqref{eq:amplidue-sqrt}. Such a square root scaling can also be derived via geometric analysis by treating the thin stiff film as a layer that cannot change its length (or area in a three-dimensional setting) but can support bending \citep{Pocivavsek2008}. By directly fitting to finite element simulations, \citet{Nguyen2020BMM} investigated the applicability of this square-root dependence to describe the evolution of surface wrinkling in a compliant HGO substrate coated with a stiff neo-Hookean film under compression. To facilitate a quantitative comparison with the post-buckling behavior shown in Figure 13 of \citet{Nguyen2020BMM}, we define the axial strain as $\epsilon = 1 - \lambda$ and denote the deflection amplitude at the top surface ($x_2 = h$) by $\mathfrak{S}$. The critical strain for surface wrinkling is denoted by $\epsilon_\mathrm{cr}$, and the corresponding wavelength is given by $\Lambda = 2\pi/n$ (note that we have fixed $n = 1$ in our analysis). \citet{Nguyen2020BMM} assume a square root dependence of $\mathfrak{S}/\Lambda=c_0\sqrt{(\epsilon-\epsilon_\mathrm{cr})/\epsilon_\mathrm{cr}}$ with $c_0$ a fitting parameter. They presented both finite element simulation results and the corresponding square root fitting curves for the evolution of $\mathfrak{S}/\Lambda$ in their Figure 13, considering the case $\mu_\mathrm{nH} \approx 100\mu_\mathrm{HGO}$ ($r \approx 0.01$), $k_1 = 2\mu_\mathrm{HGO}$ (i.e., $k_1 = 0.02$ in nondimensional form), $k_2 = 0.8393$, and three fiber orientations: $\theta = 90^\circ$, $60^\circ$, and $45^\circ$.

As discussed in previous Sections \ref{linear-bifurcation-analysis} and \ref{nonlinear-analysis}, the effect of $k_2$ is marginal.  We therefore set $k_2 = 0$ and consider the parameters $r = 0.01$, $k_1 = 0.02$, and $\theta = 90^\circ$. From the bifurcation condition \eqref{bif-exact} we find that $\epsilon_\mathrm{cr}=1-\lambda_\mathrm{cr}=0.045885$ and $(nh)_\mathrm{cr}=0.42246$. Correspondingly, the analytical post-buckling solution \eqref{am-eq} or \eqref{eq:amplidue-sqrt} yields (to avoid confusion we mention that we have used the script style $\varepsilon$ to denote the small parameter in our post-buckling analysis, cf. \eqref{increment})
\begin{equation}
   \mathfrak{S}/\Lambda=\left\{ \begin{aligned}
        &0,\hspace{32.3mm}\mbox{if }\epsilon<\epsilon_\mathrm{cr},\\
       & 0.321188\sqrt{\epsilon-\epsilon_\mathrm{cr}},\quad\mbox{if }\epsilon\geqslant\epsilon_\mathrm{cr}.
    \end{aligned}\right.
    \label{eq:com-analytical}
\end{equation}
Additionally, we determine from Figure 13 with $\theta=90^\circ$ in \citet{Nguyen2020BMM} that the fitting parameter is determined to be $c_0\approx13.5955$ such that $\mathfrak{S}/\Lambda\approx13.5955\sqrt{(\epsilon-\epsilon_\mathrm{cr})/\epsilon_\mathrm{cr}}$.

Figure \ref{fig:amplitude-strain} presents a comparison between the analytical solution \eqref{eq:com-analytical} and the square root fitting expression. As shown in Figure 13 of \citet{Nguyen2020BMM}, the fitting model accurately captures the post-buckling evolution up to the point where surface wrinkles transition to a period-doubling instability. Consequently, the good agreement observed in Figure \ref{fig:amplitude-strain} provides further validation of our analytical post-buckling solution. Although wrinkle formation begins to induce compressive stress in the fibers when the axial strain $\epsilon$ exceeds approximately $0.0562$ (indicated by the blue point in Figure \ref{fig:amplitude-strain}), the analytical solution remains accurate up to $\epsilon\approx0.12$. Beyond this strain, surface wrinkling ceases to be energetically favorable, and period doubling appears. For fiber orientations $\theta=60^\circ$ and $45^\circ$, a similarly good agreement between the analytical solution and the fitting model is observed; these results are omitted here for brevity.

\begin{figure}[!ht]
    \centering
    \includegraphics[width=4in]{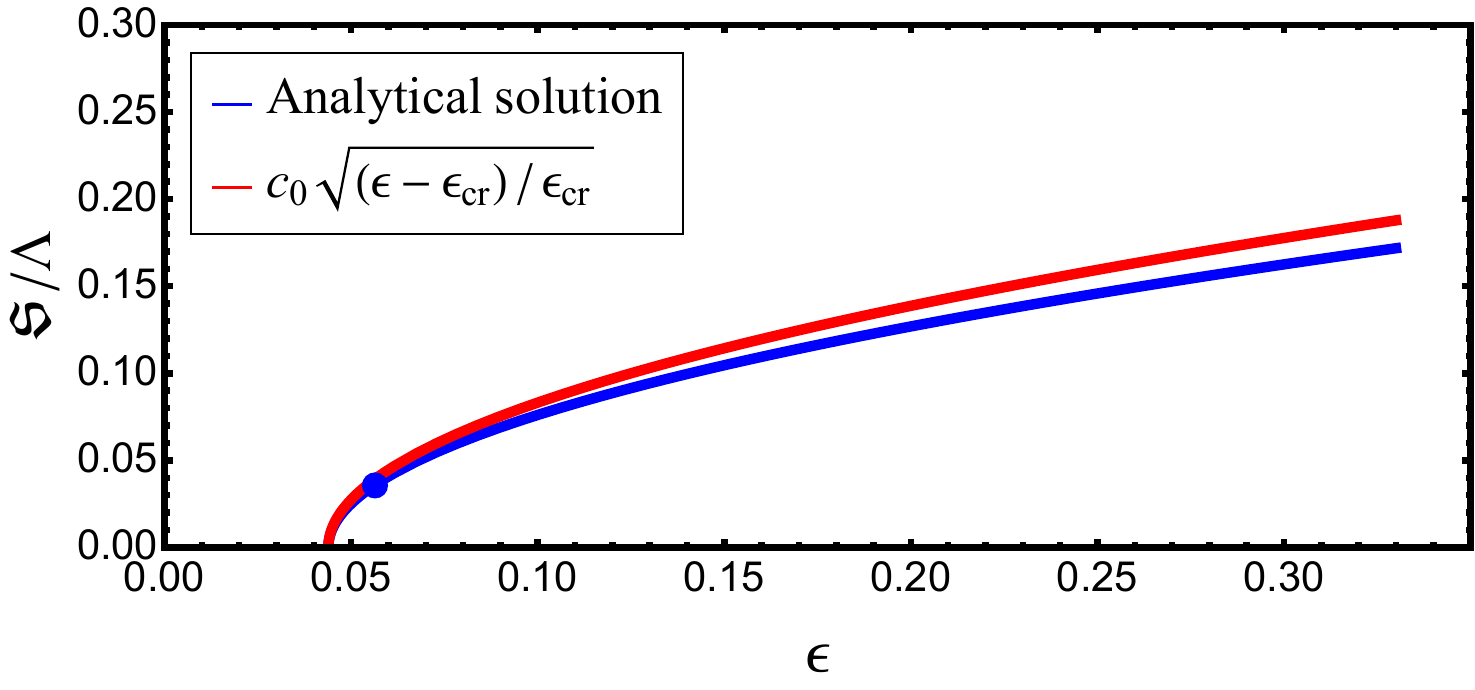}
       \caption{{\color{black}The scaled amplitude $\mathfrak{S}/\Lambda$ as a function of the applied compressive strain $\epsilon$ when $r_1=0.01$, $k_1=0.02$, $k_2=0$, and $\theta=90^\circ$. The blue curve represents the analytical solution given by \eqref{eq:com-analytical} while the red curve corresponds to the fitting result reported by \citet{Nguyen2020BMM}.}
       }
       \label{fig:amplitude-strain}
 \end{figure}

The analytical post-buckling solution not only captures the nature of the bifurcation but, when compared with finite element simulations, provides a convenient framework for understanding the influence of various geometrical and material parameters on post-buckling evolution. Moreover, it identifies the point at which the bifurcation becomes supercritical as system parameters vary. \citet{Nguyen2020BMM} also performed experiments on arterial wrinkling using pig carotid arteries to study the validity of the square root scaling for wrinkling and the linear scaling for folding, as proposed by \citet{Pocivavsek2008}. However, the fitting approach adopted in those studies does not reveal how material parameters collectively influence pattern evolution. Based on the fitting expression for post-buckling deflection shown in Figure 2 of \citet{Nguyen2020BMM}, we use our analytical solution \eqref{am-eq} to  estimate the modulus ratio $r$ between the substrate and the film. Neglecting fiber reinforcement effects ($k_1=0$), we find a good agreement with the observed wrinkling evolution  when $1/3<r<1/2$, which is consistent with the known mechanical properties of the arterial media and adventitia layers (with the thin and soft intima layer excluded from the film-substrate approximation) \citep{Giudici2021}.

Additionally, the fitting expressions are limited in their ability to provide qualitative or quantitative predictions regarding the evolution of surface wrinkling under varying geometrical or material conditions. In this work, we derive new analytical post-buckling solutions that allow for the estimation of material properties of biological tissues from measurable wrinkling patterns. Furthermore, these solutions offer a robust benchmark for validating newly developed reduced theories or numerical models for fiber-reinforced soft materials.

\subsection{When the fibers are non-symmetric}
In line with previous work by \citet{Nguyen2020BMM} and \citet{Mirandola2023JAP}, our initial model assumes a symmetric distribution of fiber families. To gain deeper insights into the role of asymmetry, we now broaden the framework to allow the two fiber families to adopt distinct preferred directions:
\begin{equation}
\mathbf{M}=\cos \theta_1 \mathbf{e}_1+\sin \theta_1 \mathbf{e}_2, \quad \mathbf{M}^{\prime}=\cos \theta_2 \mathbf{e}_1-\sin \theta_2 \mathbf{e}_2,
\end{equation}
where $\theta_1$ and $\theta_2$ is the angles between the fiber orientation and the $X_1$-axis. Using the reduced strain energy function \eqref{HGO-model-new}, the Lagrange multiplier $\bar{p}_\mathrm{HGO}$ is given by
\begin{equation}
     \bar{p}_\mathrm{HGO}=\frac{\mu_\mathrm{HGO}}{\lambda^2} +\frac{2 k_1 \sin^2 \theta_1}{\lambda^2}\left(\lambda^2 \cos^2 \theta_1-1+\frac{\sin^2 \theta_1}{\lambda^2}\right)+\frac{2 k_1 \sin^2 \theta_2}{\lambda^2}\left(\lambda^2 \cos^2 \theta_2-1+\frac{\sin^2 \theta_2}{\lambda^2}\right).
\end{equation}

It is natural to extend the general derivations presented in Sections \ref{problem-formulation}--\ref{nonlinear-analysis} to obtain both the bifurcation condition and the amplitude equation governing surface wrinkling. In this section, we briefly consider the second case where the substrate is fiber-reinforced, focusing on the influence of non-symmetric fiber distributions on the onset of surface wrinkling.

We find that the characteristic equation \eqref{cha-eq} is  replaced by 
\begin{equation}
    \beta_4 s^4+\beta_3 s^3+\beta_2s^2+\beta_1s+\beta_0=0,
    \label{eq:cha-eq-new}
\end{equation}
where $\beta_i$ $(i=0,1,\cdots,4)$ are constants related to the elastic moduli $\mathcal{A}_{jilk}$ whose expressions are omitted. When $\theta_1 = \theta_2$, the coefficients $\beta_1$ and $\beta_3$ vanish, recovering the symmetric fiber case. In general, equation \eqref{eq:cha-eq-new} admits four complex roots, two of which have positive real parts, while the other two have negative real parts. Apart from this distinction, the bifurcation condition can still be derived following the same procedure outlined in Section \ref{linear-bifurcation-analysis}.

To understand the effect of non-symmetric fibers, we fix $r=0.01$ and plot in Figure \ref{cr-theta2} the critical stretch $\lambda_\mathrm{cr}$ and the critical wavenumber $(nh)_\mathrm{cr}$ as functions of $\theta_2$, for varying $\theta_1$ and $k_1$. The angle $\theta_2$ is varied from $45^\circ$ to $135^\circ$ to avoid fiber compression. Notably, the case of a single family of fibers is recovered when $\theta_1 + \theta_2 = \pi$. For comparison, the results for the isotropic case are indicated by the black lines. We observe that the presence of non-symmetric fibers consistently delays the onset of surface wrinkling and increases the number of wrinkles, in agreement with the trends observed in Figure \ref{cr-theta-case2}. In particular, when $k_1 = 0.05$, the curves for $\theta_1 = 45^\circ$ (blue), $60^\circ$ (cyan), and $90^\circ$ (orange) exhibit similar shapes, differing primarily in phase. 

\begin{figure}[!ht]
    \centering
    {\subfigure[]{\includegraphics[width=0.35\textwidth]{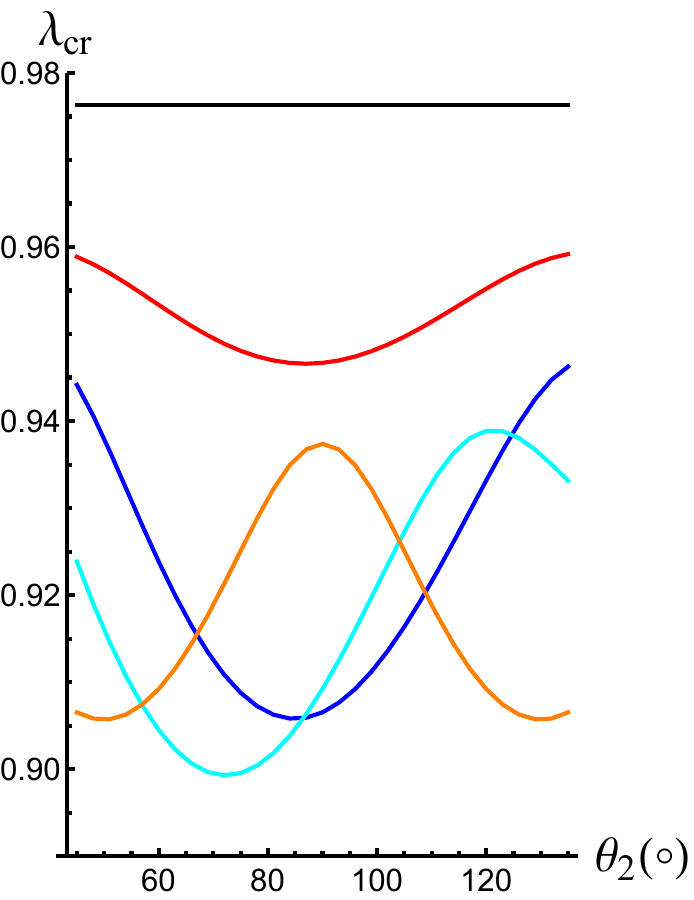}\label{lambdacr-theta2-asym}}}
    \hspace{4mm}
    {\subfigure[]{\includegraphics[width=0.55\textwidth]{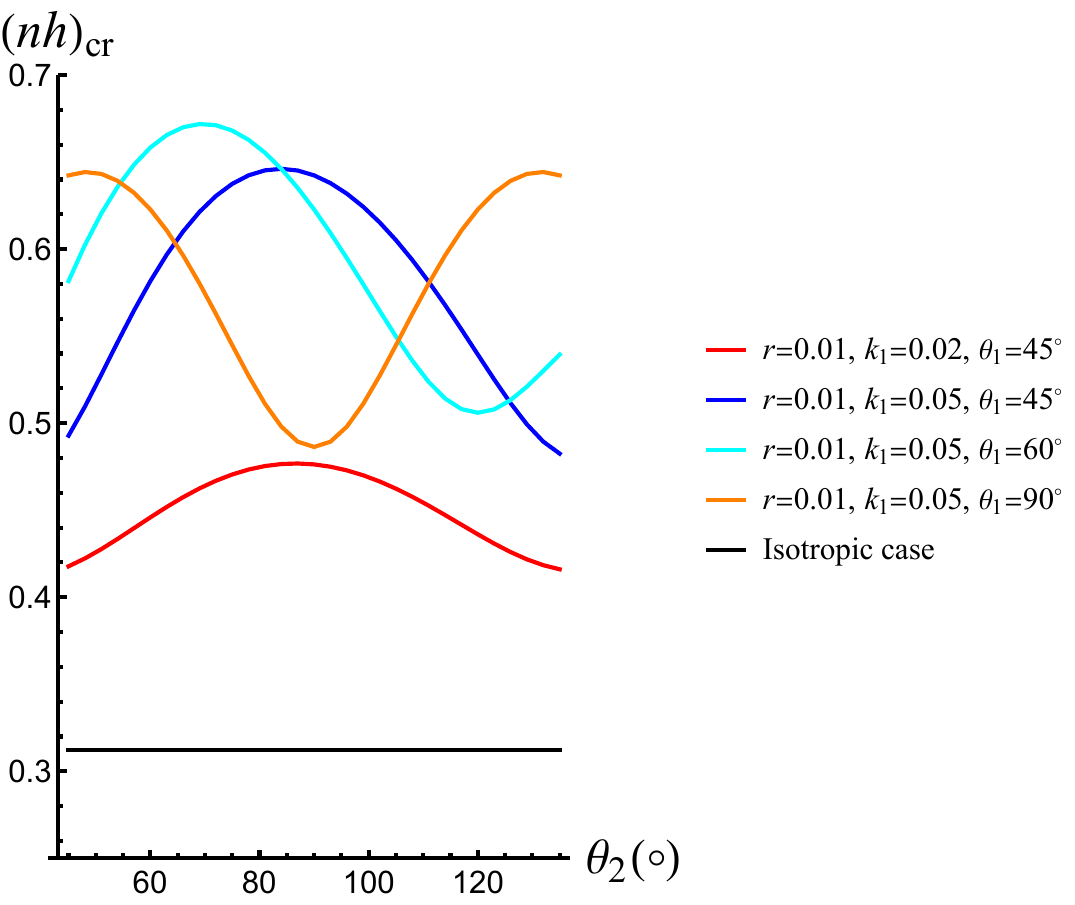}\label{mode-theta2-asym}}}
       \caption{{\color{black}The critical stretch $\lambda_\mathrm{cr}$ and the critical wavenumber $(nh)_\mathrm{cr}$ as functions of fiber angle $\theta_2$ as $r=0.01$ with different values of $k_1$ and $\theta_1$. The substrate is composed of an HGO material while the film is isotropic.}}
    \label{cr-theta2}
 \end{figure}

For symmetric fibers with $\theta_1 = \theta_2 = \theta$, it follows from equations (19) and (23) in \citet{Mirandola2023JAP} that both the critical stretch and the critical wavenumber depend on $\theta$ through a term $\sin^4(4\theta)$, which has a minimal period of $\pi/4$. This implies that $\lambda_\mathrm{cr}$ and $(nh)_\mathrm{cr}$ exhibit a quasi-periodic behavior with the same period $\pi/4$, although the peaks and valleys are not symmetric. Furthermore, this explains the non-monotonic behavior observed in all the curves in Figure \ref{cr-theta-case2}. Nevertheless, the apparent quasi-period seems to change from $\pi/4$ to $\pi/2$ in Figure \ref{cr-theta2}, suggesting a possible dependence on $\theta_2$ (or $\theta_1$) via a term like $\cos 4\theta_2$. Similarly, in the \textit{Case I} discussed in Section \ref{caseI}, the critical stretch and wavenumber shown in Figure \ref{cr-theta} also display a quasi-period of $\pi/2$. To better understand these patterns, a more comprehensive asymptotic analysis is required to obtain mathematically consistent approximations for the critical state, which we leave for a separate study.

By considering a simple non-symmetric case, we have demonstrated that asymmetry can  introduce intrinsic differences in the critical buckling condition. It is therefore expected that such asymmetry may also have a significant influence on the post-buckling behavior. This aspect will be explored in future studies.



\subsection{When the fibers can support compression}
We adopt the HGO model to describe fiber reinforcement \citep{Holzapfel2000JE,Holzapfel2010PRSA,Holzapfel2019PRSA}. A key feature of this model is that fibers are unable to support compressive loads; when subjected to compression, the fibers become inactive, and the material response reduces to that of a neo-Hookean solid. To account for this feature, a tension–compression switch is implemented to eliminate the fiber contribution under compressive strains \citep{Holzapfel2015}. This assumption is appropriate for many biological tissues, such as skin and arteries \citep{Holzapfel2006}. In view of this, we restrict our attention to fiber angles $\theta > 45^\circ$, ensuring that the fibers are under tension in the homogeneous state. This scenario is relevant to many real-world biological systems. For instance, wrinkling and folding in brain tissues often influence axonal growth, resulting in a fiber-reinforced white matter (substrate) with fiber orientations typically around $\theta \approx 90^\circ$ \citep{solhtalab2025}. We emphasize that the buckling and post-buckling results derived from the HGO model are not directly applicable when fibers are under compression. Accordingly, we make no claims  regarding the buckling behavior of bent or wrinkled fibers in this study.

In practice, fiber distributions in both biological tissues and fiber-reinforced composites can be highly complex, making the presence of compressive fibers inevitable. Moreover, when the fiber stiffness greatly exceeds that of the surrounding matrix, the fiber’s bending stiffness becomes significant and must be incorporated into the overall mechanical response \citep{Kakaletsis2023}. Additionally, fiber microbuckling plays a crucial role in shaping the macroscale buckling behavior \citep{li2018instabilities,Poulios2016}. Therefore, it is of great importance to develop theoretical models that account for the flexural and torsional elasticity of fibers in order to better understand how fiber configuration influences pattern formation and evolution of soft anisotropic materials.

Although beyond the scope of this study, we note that an effective approach to capturing the bending and twisting stiffnesses of embedded fibers is through the use of Cosserat elasticity for fiber-reinforced materials \citep{shirani2020cosserate,steigmann2023}. Recently, \citet{mcavoy2025incremental} developed an incremental theory for fibrous materials with a single family of fibers that resist extension, bending, and twisting. Their work provides a general framework for the buckling analysis of fiber-reinforced structures. As a case study, they examined compression-induced surface instability in a fibrous half-space. Unlike classical Biot instability, the presence of fibers introduces a dependence of the critical stretch on the wavenumber. Furthermore, \citet{mcavoy2025exotic} implemented Cosserat elasticity for fibrous materials within a three-dimensional finite element framework and explored novel buckling patterns induced by fiber reinforcement.

To conclude, we emphasize that the case in which fibers can support compression is considerably more complex. The application of Cosserat elasticity necessitates additional kinematic descriptors to  represent adequately fiber behavior. Nevertheless, this direction merits further investigation, as it offers the potential to yield deeper insights into the buckling and post-buckling behavior of fiber-reinforced soft tissues.
}

\section{Conclusion}\label{conclusion}

We studied surface instabilities of a fiber-reinforced bilayer  under uniaxial compression. {\color{black}With advances in 3D printing technology, it has become convenient to fabricate anisotropic materials with tailored fiber orientations \citep{Montanari2024ijss}.} Two cases were considered, i.e., either the film or the substrate is composed of a fiber-reinforced material. Within the framework of finite elasticity, the nonlinear incremental equations were formulated, from which both the linear bifurcation analysis and the weakly nonlinear analysis can be carried out. An exact bifurcation condition was obtained that explains how fiber orientation and fiber stiffness regulate surface wrinkles at onset. In addition, we obtained an amplitude equation  by a weakly nonlinear analysis. This amplitude equation describes how the wrinkled morphology grows beyond the bifurcation point and further can provide a criterion for the type of bifurcation observed. 

To capture fiber-reinforcement, we employed the Holzapfel-Gasser-Ogden model which was first proposed for arteries \citep{Holzapfel2000JE,Holzapfel2010PRSA}. Note that in this model the fibers cannot support compression. Therefore, we identified allowable values of the stretch $\lambda$ and the fiber angle $\theta$ where fibers play a role. Otherwise, when the fiber angle is below the transition angle, the anisotropic bilayer considered here will reduce to the well-known case of a neo-Hookean bilayer.

{\color{black}When the film is fiber-reinforced, wrinkling instability can be either promoted or suppressed by adjusting the fiber stiffness and orientation. In particular, increasing fiber stiffness consistently reduces the amplitude of wrinkles. To illustrate the effect of fiber orientation, we present two phase diagrams in Figure \ref{theta-k1}. Once the applied compressive strain exceeds the critical threshold, the system enters a post-buckling state. Using the amplitude equation \eqref{eq-amplitude}, we analyzed how fiber properties influence the transition modulus ratio and identify the shift between subcritical and supercritical bifurcations for various fiber angles. Additionally, we determined the admissible domain of the post-buckling solution using the HGO model and presented the corresponding bifurcation diagram.

When the substrate is fiber-reinforced, fibers can simultaneously retard or promote surface wrinkling, depending on their properties. This phenomenon parallels the process of skin wrinkling with aging, where the dermis loses stiffness due to collagen fiber degradation \citep{Kruglikov2018Skin}. A key finding was the existence of a transition fiber angle at which surface wrinkling gives way to Biot instability, characterized by a critical mode tending to infinity. To illustrate this behavior, we presented two phase diagrams in Figure \ref{fig:pha-diag}, mapping the regions of different instability modes in the $(\theta,k_1)$- and $(\theta,r)$- planes. Additionally, we identified the transition between subcritical and supercritical bifurcations and provided a bifurcation diagram derived from our post-buckling analysis. Our results show that subcritical bifurcation is  more likely to emerge when fiber stiffness dominates the substrate, offering insight into the emergence of creasing and folding patterns in biological tissues such as the brain \citep{Stewart2016EML,solhtalab2025}.

In both cases, we have validated our buckling and post-buckling results both qualitatively and quantitatively against existing results reported in \citet{Cai1999PRSA,Nguyen2020BMM,Altun2023MoSM,Mirandola2023JAP}. Specifically, the analytical post-buckling solutions derived in this study involve no \textit{ad hoc} assumptions or fitting parameters. Furthermore, the analytical solution provides a convenient way for obtaining the post-buckling evolution and understanding the influence of anisotropy on the post-buckling deflection. Additionally, it serves as a benchmark for validating reduced models and finite element schemes for fibrous materials, and provides a solid foundation for further analytical investigations into period-doubling secondary bifurcations \citep{Fu2015SIAM}.}

For $\theta>45^\circ$, fibers are expected to resist lateral deformation. Interestingly, when the film is fiber-reinforced, there exists a critical fiber angle at which the fibers become structurally undetectable, as illustrated in Figure \ref{cr-theta}. This phenomenon is analogous to the “magic angle” observed in inflated fiber-reinforced cylindrical tubes, where deformation is entirely suppressed \citep{Goriely2013PRSA}. Similarly, when the fiber orientation falls within a specific range, localized bulging in inflated tubes becomes impossible \citep{Wang2018JEM}. However, in thick fiber-reinforced tissues, where the material behaves as a substrate rather than a thin film, this special angle disappears, highlighting the highly nonlinear influence of fibers on buckling and post-buckling behavior. In addition, when bifurcation is subcritical, \citet{Shen2024JMPS} numerically demonstrated that creases emerge as a consequence of an unstable deformation path. However, the long-term evolution of these patterns remains an open question, requiring further theoretical and experimental investigations.

\bigskip
\section*{Acknowledgment}\label{Acknowledgment}
This work was supported by grants from the National Natural Science Foundation of China (Project Nos 12072227 and 12021002). Y.L. and A.G. acknowledge the UKRI Horizon Europe Guarantee MSCA (Marie Sk\l odowska-Curie Actions) Postdoctoral Fellowship (EPSRC Grant No. EP/Y030559/1). For the purpose of Open Access, the author has applied a CC BY public copyright license to any Author Accepted Manuscript (AAM) version arising from this submission.

\end{document}